\documentclass[12pt]{iopart}

\expandafter\let\csname equation*\endcsname\relax
\expandafter\let\csname endequation*\endcsname\relax

\usepackage[utf8]{inputenc}
\usepackage{iopams}
\usepackage{amsmath}
\usepackage{amsthm}
\usepackage{amssymb}
\usepackage{breqn}
\usepackage{caption}
\usepackage[british]{babel}
\usepackage{graphicx} 
\usepackage{color}
\usepackage[colorlinks=true,linkcolor=blue,citecolor=blue]{hyperref}
\usepackage{mathtools}
\usepackage[T1]{fontenc}
\usepackage{cite}
\usepackage{mathrsfs}

\begin{document}

\title{Time-dependent probability density function for partial resetting
dynamics}

\author{Costantino Di Bello$^\dagger$, Aleksei V. Chechkin$^{\dagger,\S,
\ddagger}$, Alexander K. Hartmann$^*$ Zbigniew Palmowski$^\S$, and Ralf Metzler$^{\dagger\sharp}$}
\address{$\dagger$ University of Potsdam, Institute of Physics \&
Astronomy, 14476 Potsdam, Germany\\
$\S$ Faculty of Pure and Applied Mathematics, Wroc{\l}aw University of Science
and Technology, 50-376 Wroc{\l}aw, Poland\\
$\ddagger$ Akhiezer Institute for Theoretical Physics, Kharkov 61108,
Ukraine\\
$*$ Institute of Physics, University of Oldenburg, 26129 Oldenburg, Germany\\
$\sharp$ Asia Pacific Centre for Theoretical Physics, Pohang 37673,
Republic of Korea}

\begin{abstract}
Stochastic resetting is a rapidly developing topic in the field of stochastic
processes and their applications. It denotes the occasional reset of a
diffusing particle to its starting point and effects, inter alia, optimal
first-passage times to a target. Recently the concept of partial resetting,
in which the particle is reset to a given fraction of the current value
of the process, has been established and the associated search behaviour
analysed. Here we go one step further and we develop a general technique to
determine the time-dependent probability density function (PDF) for Markov
processes with partial resetting. We obtain an exact representation of the
PDF in the case of general symmetric L{\'e}vy flights with stable index
$0<\alpha\le2$. For Cauchy and Brownian motions (i.e., $\alpha=1,2$), this
PDF can be expressed in terms of elementary functions in position
space. We also determine the stationary PDF. Our numerical analysis of the
PDF demonstrates intricate crossover behaviours as function of time.
\end{abstract}

\section{Introduction}

Stochastic processes represent a core field in non-equilibrium statistical
physics and physical chemistry \cite{landau,vankampen}. They are used as
"schematisations" \cite{levy} for systems, that are too complex to describe in
microscopic detail \cite{brenig}, and in which the dynamic of an observable
is apparently random. Stochastic processes are quite ubiquitous in nature.
Examples include, inter alia, archetypical Brownian motion \cite{haenggirev},
the passive diffusion of molecules in biological cells \cite{franosch}, animal
motion \cite{vilk1}, the motion of active particles beyond their persistence
time \cite{loewen}, tracer motion in geophysical systems \cite{brianrev},
charge carrier motion in semiconductors \cite{scher}, stock prices on
financial markets \cite{bouchaud}, or disease spreading \cite{brockmann}.

One central question in the study of stochastic processes is their ability to
locate a specific target in space \cite{benichourev}. In an unlimited space a
diffusing particle may significantly stray away from its starting point and
may not be able to locate a small target in a finite interval of time. Even
in a finite domain the diffusive search may have very broad distributions of
search times, and the typical search time may be significantly different from
the mean \cite{carlos,aljaz,denis}. Speedup of the diffusive search may, e.g.,
be achieved by "facilitated diffusion", in which the diffusion intermittently
occurs in the embedding space and on a surface with reduced dimension---a
prominent example is the search of binding proteins for a site on a long
DNA chain \cite{bvh,gijs}. The central idea in facilitated diffusion is the
combination of thorough local search and decorrelations by bulk diffusion
\cite{adam,mirnyjpa}. Similar principles in random search are processes
with long-tailed jump length distributions (L{\'e}vy flights and walks)
\cite{ghandibook,sims,vladpnas,vladjstat,vladjpa,dybiecjpa,vladepjb,vladnjp,
amin,mich} and intermittent search \cite{benichourev,olivier,olivier1,heiko}.

Another way to optimise the search for a target at a finite distance away
from where the searching particle is released, is stochastic resetting (SR)
\cite{masaprl,masajpa}. In its simplest version, SR considers a Brownian
particle, that experiences repeated restarts, i.e., resets to its starting
position, either at fixed periods or stochastically with a fixed rate
\cite{masaprl,masajpa,besga,pal,bhat}. A central feature of SR is that the
stochastic search of a diffusing particle for a target at a given distance
from its starting point can be optimised for a specific resetting frequency
\cite{masaprl,masajpa,besga}. The idea is that SR prevents long departures of
the particle away from its target. Overly frequent resetting, in contrast,
keeps the particle always close to the starting point, such that it cannot
reach the target. At intermediate resetting frequencies, therefore, the mean
search time is minimised \cite{masaprl,masajpa,besga}. For mean search times
a unified approach allows to determine the optimal SR-rate \cite{sr1} and, at
optimality, first-passage time fluctuations have a universal coefficient of
variation \cite{sr2}, see also recent results on extremes in SR \cite{max}.
SR leads to a non-equilibrium steady state with a well-defined limiting
displacement distribution \cite{masaprl,masajpa,besga}. A renewal approach
to resetting was established and exploited to show that constant pace
resetting minimises the mean hitting time \cite{cheso}. Moreover, linear
response and fluctuation-dissipation relations for SR were discussed
\cite{igorlr}. Aspects of SR in quantum walks have also been addressed
\cite{quant}. A recent review of SR and applications in different disciplines
can be found in \cite{martinrev}. Importantly, we mention that the effect
of SR was demonstrated experimentally \cite{exp,exp1,exp2}.

Various aspects beyond Brownian SR have been discussed. Inter alia,
non-instantaneous returns \cite{anna2} and soft resetting by switching
harmonic potentials \cite{pengbo} were studied. SR of anomalous
diffusion processes include heterogeneous diffusion processes with
distance-dependent diffusion coefficient \cite{andrey,trifce1}, scaled
Brownian motion with time-dependent diffusivity in renewal and non-renewal
settings \cite{anna, anna1}, and continuous time random walk processes with
complete and incomplete \cite{igorjpa} as well as with power-law \cite{anna3}
resetting.  Reset rotational motion was studied in terms of a time-fractional
Fokker-Planck equation \cite{irina}. Different effects due to resetting were
demonstrated for geometric Brownian motion without \cite{gbm} and with drift
\cite{gbm1}, and effects on income dynamics explored \cite{trifce}. Aspects
of ergodicity restoration in anomalous diffusion processes were also analysed
\cite{wei}. For SR on networks \cite{net,net1,net2}, the minimisation of
global mean first passage times for specific centrality-based SR mechanisms
were reported \cite{kiril}. We note that results similar to SR for a single
absorbing target were obtained for multiple as well as partially absorbing
targets \cite{bressloff,bressloff1}.  Moreover, a concept similar to SR
is preferential relocations, which take the walker back to any previously
visited site \cite{vilk,boyer}.

Here we address the question as to what happens when the particle is not reset
to its origin, but to some value in between the instantaneous co-ordinate
and the initial value. Such \emph{partial stochastic resetting} (PSR) has
been studied in mathematical \cite{dumas, lopker}, financial and actuarial
\cite{ruinprobs, marciniak, boxma,boxma1,hofstad} literature, and in queueing
theory \cite{lopkertcp} for piecewise deterministic processes. The basic idea
behind many models in these fields is that there is a growing observable
(like the income of an insurance company or the amount of traffic over the
internet), subjected to random unexpected events leading to a substantial
decrease of this quantity (claims in an insurance company or failures in
internet connections). PSR has also been recently considered in physics
literature \cite{marcus}, where the authors studied the two distinct cases
of independent and dependent random resetting amplitudes: for independent
resetting, the amplitude is arbitrary, so that the particle can also be reset
to negative values, while for dependent resetting amplitudes the current
value of the particle is multiplied by a number between zero and unity,
thus guaranteeing positivity of the value after reset. In \cite{marcus}
the authors discussed PSR for both scenarios in terms of moments and the particle
probability density function (PDF). The case of dependent resetting
was recently also analysed further \cite{pierce,shlomi}. PSR finds its
motivation in different settings. One is stratigraphy, studying sediment
layering in geology \cite{marcus}: deposits by a gradual sedimentation,
e.g., in a river delta, can be partially washed away by sudden events such
as extreme rainfall. A similar model is used in population dynamics, when
the gradual growth dynamic is interrupted by sudden, catastrophic population
decimation \cite{gripenberg,pakes,brockwell,hanson, artalejo}.

Going significantly beyond recent work \cite{pierce,shlomi} reporting the
Fokker-Planck equation and the stationary PDF for Brownian PSR \cite{pierce}
and the time-dependent PDF in Fourier-Laplace space when the initial
condition is at the origin, we here develop a general technique to determine
the time-dependent PDF for homogeneous Markov processes with
Poissonian resetting, in which the process is partially reset by
multiplication with the constant factor $0<c<1$ at random times $T_1,T_1+T_2,
\ldots$. The limiting cases $c\to0$ and $c\to1$ of this model correspond to
total resetting \cite{masaprl} and a stochastic process without resetting,
respectively. An exact representation of the PDF in the real space-time
domain is derived for the case of general symmetric L{\'e}vy flights with
stable index $0<\alpha\le2$, including Brownian motion and Cauchy flights
as particular cases for $\alpha=2$ and $1$, respectively. We also determine
the stationary PDF for symmetric L{\'e}vy flights in terms of Fox $H$-functions
and present the particular cases $\alpha=2$ and $1$ in terms of elementary
functions. For the case of non-zero initial conditions, we report highly
asymmetric non-stationary PDFs for $\alpha=2$ and the emergence of non-trivial
inhomogeneous multimodal regimes with $\alpha\neq2$.

\section{Propagator for partial resetting}
\label{section:general_derivation}

We consider a stochastic process $X_t$ with initial condition $X_0=x_0$
whose PDF is $p_0(x, t|x_0, t_0)$. We assume homogeneity in both space and
time, such that $p_0(x,t|x_0,t_0)=p_0(x-x_0,t-t_0|0,0)$. Without limitation of
generality, we set $t_0=0$ and use the simplified notation $p_0(x,t)$, keeping
the initial condition $x_0$ implicit. At random times $T_1,T_1+T_2,T_1+T_2+T_3,
\ldots$, the position of the particle is partially reset, i.e., multiplied
by $0<c<1$. Thus $T_i$ represents the time between the $(i-1)$st and $i$th
partial reset. We assume that the $T_i$ are independent, identically
distributed (i.i.d.) random variables with PDF $\psi(t)=\mathrm{Pr}\{
t\leq T_1\leq t+dt\}$. Clearly, $c=0$ represents the full resetting case,
while setting $c=1$ we retrieve the unperturbed stochastic process. Let $Y_t$
denote the PSR process. We then have
\begin{eqnarray}
\nonumber
Y_t&=&x_0+\left[\left(\left( X_{T_1}\cdot c +X_{T_2}\right)\cdot c +X_{T_3}\right)\cdot c +\dots \right]\cdot c +X_{t-T_{N_t}} \\
&=&x_0 + c^{N_t}X_{T_1} + c^{N_t-1}X_{T_2} +\dots cX_{T_{N_t}} +X_{t-T_{N_t}},
\label{psr}
\end{eqnarray}
where $N_t$ denotes the number of partial resetting events in the time interval
$[0,t]$. The meaning of this expression is quite intuitive: the process is
unperturbed until the time $T_1$, moving from $x_0$ to $x_0+X_{T_1}$; then
the process is multiplied by $c$, and it stays unperturbed again between
times $T_1$ and $T_1+T_2$, and so on. Some possible trajectories of $Y_t$
are depicted in figure for Brownian and Cauchy random walks (see below) in
absence and presence of PSR \ref{fig:trajs}. Generally we notice that in the
presence of PSR the resulting trajectories tend to be closer to the origin,
while they experience long excursions in the unperturbed case. This hints at
the existence of a stationary state, that we will examine more closely below.

\begin{figure}
\centering
\includegraphics[scale=0.35]{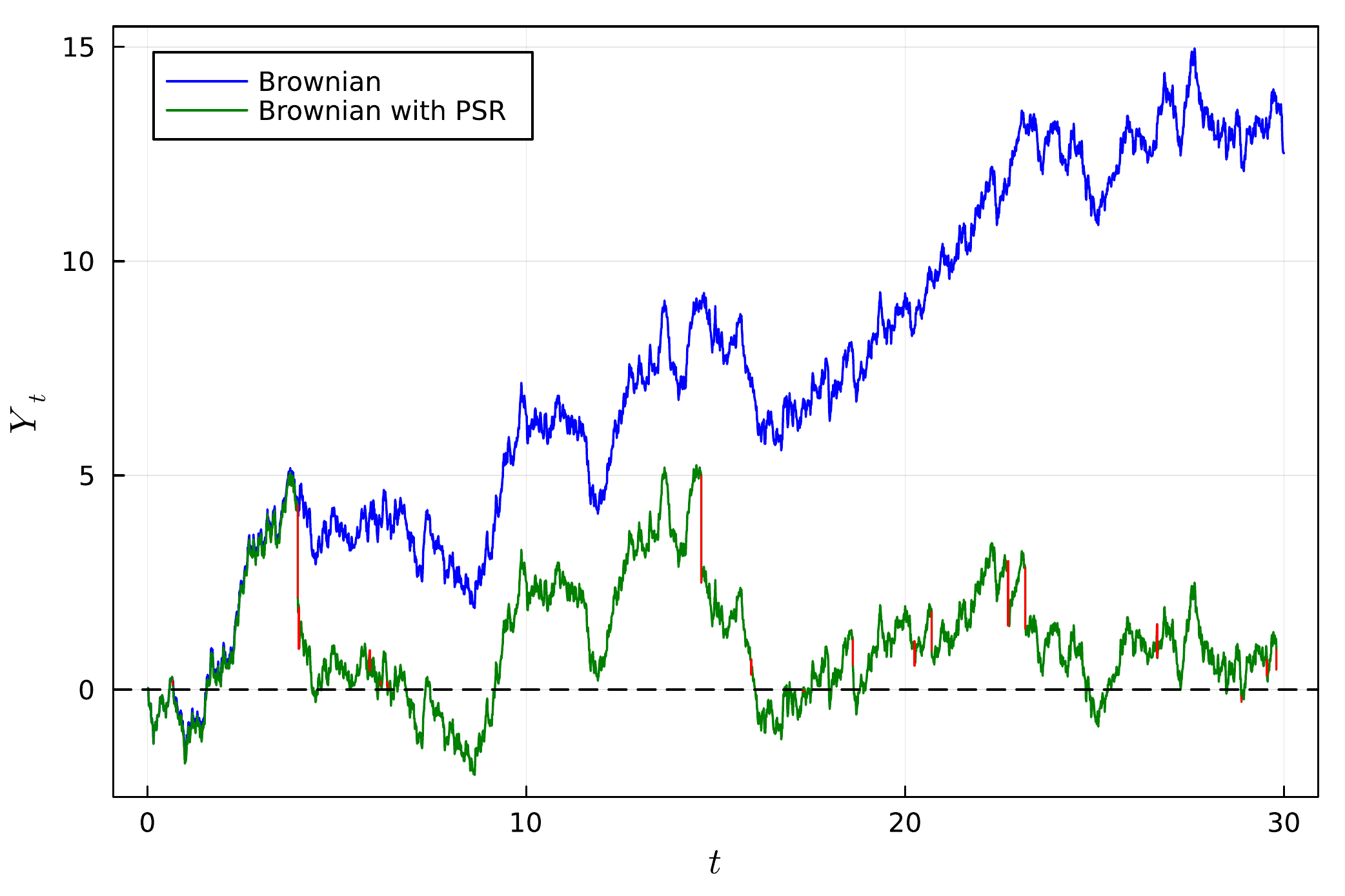}
\includegraphics[scale=0.35]{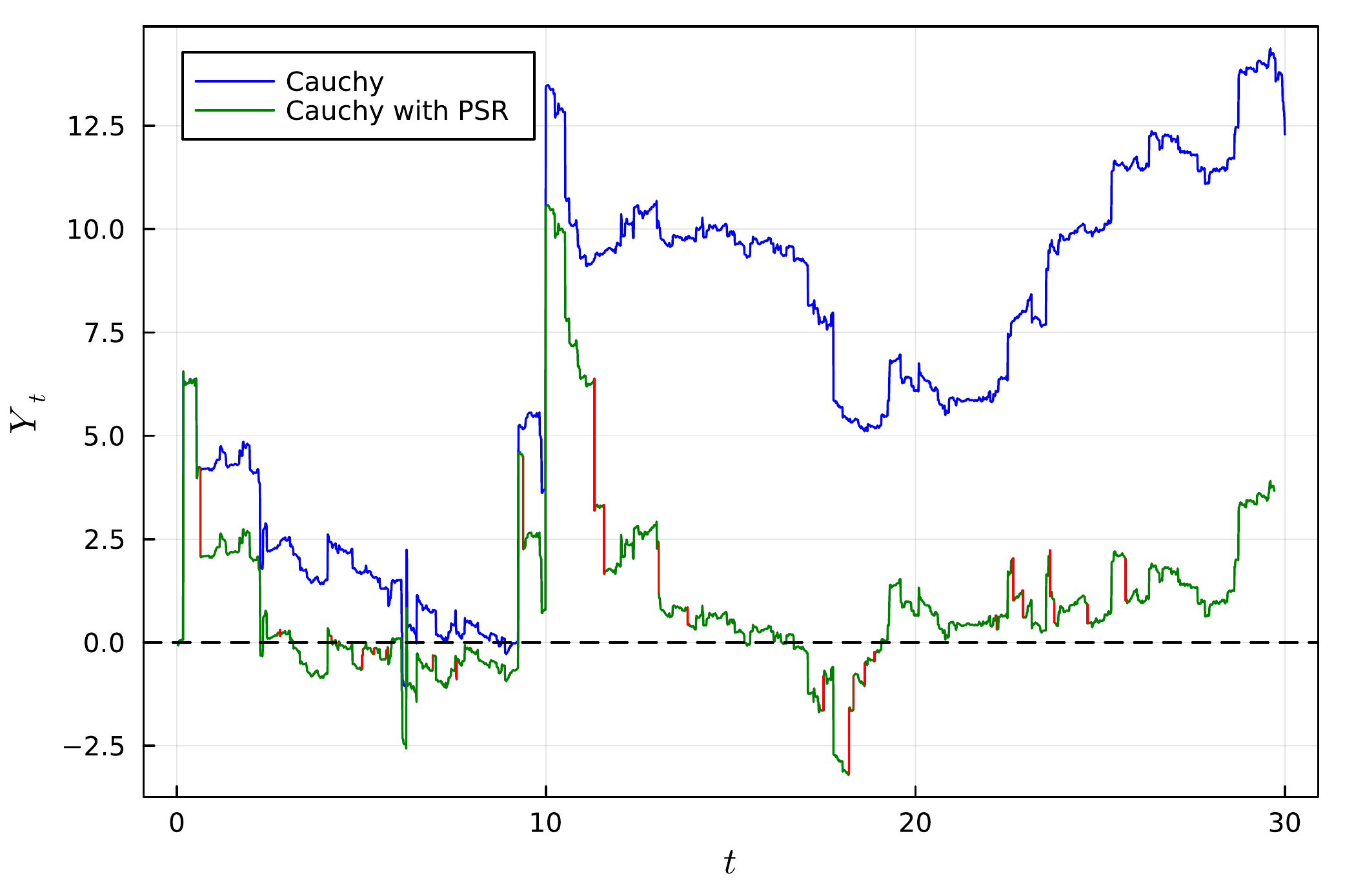}
\caption{Plot of two possible trajectories of the system in the specific
cases of Brownian motion (left panel) and symmetric Cauchy random walk (right
panel). The process without PSR ($X_t$) is depicted in blue on both panels,
while the process with PSR ($Y_t$, with rate $r=1.0$ and amplitude factor
$c=0.5$) is displayed in green. We highlighted partial resetting events with
red segments. The processes $X_t$ and $Y_t$ in either panel were generated
from the same random seed.}
\label{fig:trajs}
\end{figure}

We are interested in finding the PDF $p_r(x,t|x_0)$ of the PSR process $Y_t$.
Since we choose $X_t$ to be time-homogeneous, it follows that $Y_t$ is also
homogeneous in time. However, partial resetting according to equation
(\ref{psr}) leads to an inhomogeneity in space. Thus, in the PDF $p_r(x,t|x_0)$
we removed the dependence on the initial time $t_0$ (taken as $t_0=0$) but we
retain the dependence on $x_0$. The reason for the loss of spatial homogeneity
is quite intuitive: consider the first partial resetting event, occurring at
time $T_1$. The position $Y_{T_1^+}$ depends on $Y_{T_1^-}$, and $Y_{T_1^-}$
in turn depends on $x_0$. Due to partial resetting the shape of the resulting
time-dependent PDF due to this effect attains more complicated shapes, as we
will discuss in the next sections.

For the specific case of Poissonian resetting times, i.e., $\psi(t)=re^{-rt}$
for all $i$, the expression of $p_r$ can be found through the last renewal
equation, which reads
\begin{equation}
\label{eqn:last_renewal_equation}
p_r(x,t|x_0)=e^{-rt}p_0(x,t|x_0)+\int_0^tdt're^{-rt'}\int_{-\infty}^{\infty}dy
p_r(y,t-t'|x_0)p_0(x,t'|cy).
\end{equation}
The meaning of this relation is quite simple: the first term on the right hand
side takes into consideration all realisations in which no partial resetting
occurred, while the second term considers all realisations in which the last
resetting event occurred at time $t-t'$. During the time interval $[0,t-t']$
the particle diffuses to position $y$ with propagator $p_r(y,t-t'|x_0)$, while
during $[t-t', t]$ it diffuses without PSR, hence with $p_0(x,t'|cy)$. This
term must be integrated over all possible realisations of $t'$ and $y$. The
solution of the integral equation \eqref{eqn:last_renewal_equation} can be
obtained via the series expansion
\begin{equation}
\label{eqn:series_expansion_pr}
p_r(x,t|x_0)=e^{-rt}\sum_{n=0}^{\infty}r^nq_n(x,t |x_0),
\end{equation}
where the set of functions $\{ q_n\}_{n=0}^{\infty}$ can be found through the recursion relation
\begin{equation}
\label{eqn:qn_definition}
q_0(x,t|x_0)=p_0(x,t|x_0),\quad
q_n(x,t|x_0)=\int_0^tdt'\int_{-\infty}^{\infty}dyq_{n-1}(y,t'|x_0)p_0(x,t-t'|cy).
\end{equation}
We proof this result in \ref{appendix:last_renewal}. In the recursion relation
(\ref{eqn:qn_definition}) we now perform a Laplace transform to obtain\footnote{We
use the notation
\[
\mathscr{L}\{f(t)\}(s)=\int_0^{\infty}e^{-st}f(t)dt=\tilde{f}(s).
\]}
\begin{equation}
\tilde{q}_n(x,s|x_0)=\int_{-\infty}^{\infty}dy\tilde{q}_{n-1}(y,s|x_0)\tilde{p}
_0(x-cy,s).
\end{equation}
Note that the explicit dependence on $x_0$ is also inherent in these transformed
functions. Applying an additional Fourier transform,\footnote{The Fourier
transform is defined as
\[
\mathscr{F}\{g(x)\}(k)=\int_{-\infty}^{\infty}e^{ikx}g(x)dx=\hat{g}(k).
\]
}
\begin{equation}
\label{eqn:double_transform_qn}
\hat{\tilde{q}}_n(k,s|x_0)=\hat{\tilde{q}}_{n-1}(kc,s|x_0)\hat{\tilde{p}}_0(k,s).
\end{equation}
Thus we can find the general expression by simply iterating
\begin{equation}
\hat{\tilde{q}}_n(k,s|x_0)=\left(\prod_{l=0}^{n-1}\hat{\tilde{p}}_0(kc^l,s)\right)
\hat{\tilde{p}}_0(kc^n,s|x_0).
\end{equation}
Since $p_0$ is spatial homogeneous we can use the relation $\hat{\tilde{p}}_0(k
c^n,s|x_0)=e^{ikc^nx_0}\hat{\tilde{p}}_0(kc^n,s)$ to obtain
\begin{equation}
\label{eqn:qn_double_transform}
\hat{\tilde{q}}_n(k,s|x_0)=e^{ikc^nx_0}\prod_{l=0}^{n}\hat{\tilde{p}}_0(kc^l,s).
\end{equation}
Hence, combining \eqref{eqn:series_expansion_pr} and
\eqref{eqn:qn_double_transform} we may write the full PDF in Fourier-Laplace
space as
\begin{equation}
\label{eqn:propagator_general_result}
\hat{\tilde{p}}_r(k,s|x_0)=\sum_{n=0}^{\infty}r^ne^{ikc^nx_0}\prod_{l=0}^n
\hat{\tilde{p}}_0(kc^l,r+s).
\end{equation}
Equation \eqref{eqn:propagator_general_result} is the first main result of
the paper, generalising previous results \cite{marcus,shlomi,pierce}. In
\cite{marcus} the authors considered the case of deterministic ballistic
motion with constant speed and PSR. The propagator for this process which
was not reported in \cite{marcus} can be found from result
(\ref{eqn:propagator_general_result}) by setting $p_0(x,t)=\delta(x-vt)$,
where $v$ is the speed. Moreover, the results in \cite{shlomi,pierce}
follow from \eqref{eqn:propagator_general_result} by setting
$x_0=0$ and $p_0(x,t)=(4\pi Dt)^{-1/2}\exp(-x^2/(4Dt))$.

For consistency, we check the limit of $c\to1$, for which we find
\begin{eqnarray}
\nonumber
\hat{\tilde{p}}_r(k,s|x_0)&=&e^{ikx_0}\hat{\tilde{p}}_0(k,r+s)\sum_{n=0}^{
\infty}\left(r\hat{\tilde{p}}_0(k,r+s)\right)^n
=e^{ikx_0}\hat{\tilde{p}}_0(k,r+s)\dfrac{1}{1-r\hat{\tilde{p}}_0(k,r+s)}\\
&=&e^{ikx_0}\hat{\tilde{p}}_0(k,s),
\end{eqnarray}
where in the last step we used an identity for Markov processes proved in
\ref{appendix:Fourier_Laplace}. As expected, we retrieve the PDF for the
stochastic process without resetting. In the case $c\to 0$, we obtain
\begin{equation}
\hat{\tilde{p}}_r(k,s|x_0)=\hat{\tilde{p}}_0(k,r+s)\sum_{n=0}^{\infty}\left(
\dfrac{r}{r+s}\right)^n=\dfrac{r+s}{s}\hat{\tilde{p}}_0(k,r+s),
\end{equation}
which is the same result as the one for total resetting \cite{martinrev}.

\section{L{\'e}vy flights}

Let us consider now the general case in which the underlying process $X_t$ is
a symmetric L{\'e}vy flight \cite{bouchaud,pccp}. The associated characteristic
function $\hat{p}_0(k,t)$ of a symmetric L{\'e}vy stable PDF is then given by
\cite{pccp,hughes,bouchaud1,sato,ourrev}
\begin{equation}
\hat{p}_0(k,t)=e^{-D|k|^\alpha t},
\end{equation}
which in Fourier-Laplace space reads (see also \cite{Linnik})
\begin{equation}
\label{eqn:hat_tilde_p0}
\hat{\tilde{p}}_0(k,s)=\dfrac{1}{s+D|k|^\alpha}.
\end{equation}
In real space this PDF becomes
\begin{equation}
\label{eqn:p0_integral_form}
p_0(x,t)=\int_0^{\infty}\dfrac{dk}{\pi}\cos\left(kx\right)e^{-D|k|^\alpha t},
\end{equation}
with $\alpha\in(0,2]$. The case $\alpha=2$ corresponds to a Gaussian PDF, while
for $\alpha\in(0,2)$ the asymptotic scaling of the PDF has the power-law tails
$p_0(x,t)\simeq|x|^{-1-\alpha}$ \cite{pccp,sato,ourrev,hughes,bouchaud1}. The
inverse
Fourier transform in (\ref{eqn:p0_integral_form}) can be performed by use of
Fox $H$-functions (see below), while in the special cases $\alpha=1,2$ simple,
explicit forms for the PDF $p_0(x,t)$ can be found in terms of a Cauchy PDF and
a normal Gaussian, respectively. We will treat these two special cases in detail
in the next sections.

In Fourier-Laplace space, using equations \eqref{eqn:qn_double_transform} and
\eqref{eqn:hat_tilde_p0} we obtain the functions
\begin{equation}
\label{eqn:hat_tilde_qn_Levy_flights}
\hat{\tilde{q}}_n(k,s|x_0)=e^{ikc^nx_0}\prod_{l=0}^n\dfrac{1}{s+Dc^{\alpha
l}|k|^\alpha},
\end{equation}
and with equation \eqref{eqn:propagator_general_result} we find
\begin{equation}
\label{eqn:hat_tilde_pr_Levy_flights}
\hat{\tilde{p}}_r(k,s|x_0)=\sum_{n=0}^{\infty}r^n e^{ikc^nx_0}\prod_{l=0}^n
\dfrac{1}{r+s+Dc^{\alpha l}|k|^\alpha}.
\end{equation}
This PDF solves the fractional Fokker-Planck equation\footnote{This differential
equation is sometimes called "pantograph" form, where this term means that there
are multiple points as arguments of the functions, in this case $x$ and $\frac{
x}{c}$, compare \cite{pantograph}.} (as shown in \ref{appendix:equivalence}) \begin{equation}
\label{eqn:Fokker-Planck_equation}
\frac{\partial p_r(x,t|x_0)}{\partial t}=D\frac{\partial^\alpha}{\partial
|x|^\alpha}p_r(x,t|x_0)-rp_r(x,t|x_0)+\frac{r}{c}p_r\left(\frac{x}{c},t|x_0
\right),
\end{equation}
where the space-fractional operator is defined in terms of its Fourier
transform, $\mathscr{F}\{\partial^{\alpha}g(x)/\partial|x|^{\alpha}\}=-|k|^{
\alpha}g(k)$ \cite{report}. Setting $\alpha=2$ and $x_0=0$ we retrieve the
dynamic equation obtained in \cite{shlomi} corresponding to Brownian motion
with PSR, see also below. For $0<c<1$ we can simplify equation
\eqref{eqn:hat_tilde_qn_Levy_flights} by using the partial fraction
decomposition\footnote{Suppose having a function $f:z\in\mathbb{C}\to\mathbb{C}$
having $n$ poles $z_1,z_2,\dots,z_n$ of order $1$. Then it holds that
\[
f(z)=2\pi i\sum_{i=1}^n\dfrac{1}{z-z_i}\text{Res}\left(f,z_i\right),
\]
where $\text{Res}(f,z_i)$ denotes the residue of the function at the pole $z_i$
\cite{ahlfors}.}
\begin{equation}
\hat{\tilde{q}}_n(k,s|x_0)=\dfrac{e^{ikc^nx_0}}{s^n}\sum_{m=0}^n\dfrac{1}{(c^{-
\alpha};c^{-\alpha})_m(c^\alpha;c^\alpha)_{n-m}}\dfrac{1}{s+Dc^{\alpha m}
|k|^{\alpha}},
\end{equation}
where the symbols in the parentheses denote the $q$-Pochhammer symbol defined
as \cite{special_functions}
\begin{equation}
(a;q)_n=\prod_{l=0}^{n-1}(1-aq^l).
\end{equation}
After inverse Laplace transform of $\hat{\tilde{q}}_n$ we obtain
\begin{equation}
\hat{q}_0(k,t|x_0)=e^{ikc^n x_0-D|k|^\alpha t},
\end{equation}
for $n=0$ and, by use of the convolution theorem,
\begin{equation}
\hat{q}_n(k,t|x_0)=e^{ikc^nx_0}\sum_{m=0}^n\dfrac{1}{(c^{-\alpha};c^{-\alpha})_m
(c^\alpha;c^\alpha)_{n-m}}\int_0^tdt'\dfrac{(t-t')^{n-1}}{(n-1)!}e^{-Dc^{\alpha
m}|k|^\alpha t'}
\end{equation}
for $n\geq 1$. We now perform an inverse Fourier transform,
\begin{equation}
q_n(x,t|x_0)=\sum_{m=0}^n\dfrac{1}{(c^{-\alpha};c^{-\alpha})_m (c^\alpha;c^\alpha)
_{n-m}}\int_0^tdt'\dfrac{(t-t')^{n-1}}{(n-1)!}p_0(x-c^n x_0,c^{\alpha m} t'),
\end{equation}
for $ n\geq1$. Then the formula for the propagator may be written in the compact
form
\begin{eqnarray}
\nonumber
p_r(x,t|x_0)&=&e^{-rt}\sum_{n=0}^{\infty}r^n\sum_{m=0}^n\dfrac{1}{(c^{-\alpha};
c^{-\alpha})_m(c^\alpha;c^\alpha)_{n-m}}\\
&&\times\int_0^tdt'\left((1-\delta_{n0})\dfrac{(t-t')^{n-1}}{(n-1)!} +\delta_{n0}
\delta(t-t')\right)p_0(x-c^n x_0,c^{\alpha m}t'),
\label{eqn:pr_Levy_flight}
\end{eqnarray}
where $\delta_{ij}$ denotes the Kronecker delta and $\delta(t)$ denotes the Dirac
$\delta$-function. The formula above is the second main result of the paper.

Let us show that expression (\ref{eqn:pr_Levy_flight}) is indeed normalised. To
this end we integrate over $x$. Since $p_0$ is normalised, we get
\begin{equation}
\int_{-\infty}^{\infty}dx p_r(x,t|x_0)=e^{-rt}\sum_{n=0}^{\infty}\dfrac{(rt)^n}{
n!}\sum_{m=0}^n\dfrac{1}{(c^{-\alpha};c^{-\alpha})_m(c^\alpha;c^\alpha)_{n-m}}. 
\end{equation}
We prove in \ref{appendix:identity} that
\begin{equation}
\label{identity}
\sum_{m=0}^n\dfrac{1}{(c^{-\alpha};c^{-\alpha})_m(c^\alpha;c^\alpha)_{n-m}}=1,
\end{equation}
and therefore $p_r(x,t|x_0)$ in (\ref{eqn:pr_Levy_flight}) is normalised, as it
should be.

\subsection{Stationary distribution}

The stationary distribution for L{\'e}vy flight-PSR can be obtained by setting
the time derivative in \eqref{eqn:Fokker-Planck_equation} to $0$ and applying
an inverse Fourier transform,
\begin{equation}
-D|k|^\alpha\hat{p}^{(s)}_r(k)-r\hat{p}^{(s)}_r(k)+r\hat{p}^{(s)}_r(kc)=0,
\end{equation}
which after iteration produces
\begin{equation}
\label{eqn:stat_distrib_general_Fourier}
\hat{p}^{(s)}_r(k)=\prod_{l=0}^{\infty}\dfrac{r}{r+Dc^{\alpha l}|k|^\alpha}.
\end{equation}
In the limiting case $\alpha=2$ we obtain the same result as in \cite{shlomi,
pierce}. We may ask whether by taking the limit $t\to \infty$ in the general
expression \eqref{eqn:hat_tilde_pr_Levy_flights} for the propagator we get
the same formula \eqref{eqn:stat_distrib_general_Fourier}. This agreement
can indeed be demonstrated, compare the use of Cesaro's and final value
theorems for the specific case $\alpha=2$ as shown in one of the appendices
of \cite{shlomi}. The derivation can be directly extended for the generic
$\alpha$. Equation \eqref{eqn:stat_distrib_general_Fourier} can be
transformed by using partial fraction decomposition, yielding
\begin{equation}
\label{eqn:hat_levy_stat}
\hat{p}^{(s)}_r(k)=\sum_{n=0}^{\infty}\dfrac{r}{r+Dc^{\alpha n}|k|^{\alpha}}
\prod_{\substack{l=0,l\neq n}}^{\infty}\dfrac{1}{1-c^{\alpha(l-n)}}
=\dfrac{1}{(c^\alpha;c^\alpha)_{\infty}}\sum_{n=0}^{\infty}\dfrac{1}{(c^{-
\alpha};c^{-\alpha})_n}\dfrac{r}{r+Dc^{\alpha n}|k|^{\alpha}}, 
\end{equation}
which is the third main result of the paper. We note that for $k=0$, by using
a well-known identity for $q$-Pochhammer symbols first discovered by Euler
\cite{special_functions},
\begin{equation}
\dfrac{1}{(c^\alpha;c^\alpha)_{\infty}}\sum_{n=0}^{\infty}\dfrac{1}{(c^{-\alpha};
c^{-\alpha})_n}=1,
\end{equation}
we see that the PDF is normalised. After inverse Fourier transform in equation
\eqref{eqn:hat_levy_stat} we obtain the stationary PDF in position space,
\begin{equation}
\label{eqn:stat_distrib_general}
p^{(s)}_r(x)=\dfrac{1}{(c^\alpha;c^\alpha)_{\infty}} \sum_{n=0}^{\infty}
\dfrac{1}{(c^{-\alpha}; c^{-\alpha})_n}\int_0^{\infty}\dfrac{dk}{\pi}\cos(kx)
\dfrac{r}{r+Dc^{\alpha n}|k|^{\alpha}}.
\end{equation}
This is another central result of this paper.

In the last expression the Fourier cosine integral can be solved analytically
in terms of Fox $H$-functions \cite{mathai}. To this end we note first that the
image function can be identified with the $H$-function
\begin{equation}
\frac{1}{1+\frac{D}{r}c^{\alpha n}|k|^{\alpha}}=H^{1,1}_{1,1}\left[\frac{D}{r}
c^{\alpha n}|k|^{\alpha}\left|\begin{array}{ll}(0,1)\\(0,1)\end{array}\right.
\right]=\frac{1}{\alpha}H^{1,1}_{1,1}\left[\left(\frac{D}{r}\right)^{1/\alpha}
c^{n}|k|\left|\begin{array}{ll}(0,1/\alpha)\\(0,1/\alpha)\end{array}\right.\right],
\end{equation}
where in the second step we made use of a well known theorem of $H$-functions
\cite{mathai}. The cosine transform then is merely a manipulation of indices
\cite{walterg}, and we find
\begin{equation}
\label{eqn:stat_Fox}
p^{(s)}_r(x)=\dfrac{1}{(c^\alpha;c^\alpha)_{\infty}}\sum_{n=0}^{\infty}\dfrac{1}{
(c^{-\alpha};c^{-\alpha})_n}\frac{1}{\alpha|x|}H^{2,1}_{2,3}\left[\frac{\lambda^{
1/\alpha}|x|}{c^{n}}\left|\begin{array}{ll}(1,1/\alpha),(1,1/2)\\(1,1),(1,1/
\alpha),(1,1/2)\end{array}\right.\right],
\end{equation}
another main result of this work. Here we defined $\lambda=r/D$. We will consider
in detail the cases $\alpha=1,2$ in the following sections.

\section{Brownian motion with PSR}

In the Gaussian case $\alpha=2$ the PDF $p_0$ reads
\begin{equation}
p_0(x,t)=\dfrac{1}{\sqrt{4\pi Dt}}\exp\left(-\dfrac{x^2}{4Dt}\right), 
\end{equation}
hence the propagator \eqref{eqn:pr_Levy_flight} becomes
\begin{eqnarray}
\nonumber
p_r(x,t|x_0)&=&e^{-rt}\frac{1}{\sqrt{4\pi Dt}}\exp\left(-\dfrac{(x-x_0)^2}{4Dt}
\right)+re^{-rt}\sum_{n=1}^{\infty}\int_0^tdt'\frac{[r(t-t')]^{n-1}}{(n-1)!}\\
&&\times\sum_{m
=0}^n\mathcal{C}^{(2)}_{n,m}\frac{1}{\sqrt{4\pi Dc^{2m}t'}}\exp\left(-\frac{
(x-c^n x_0)^2}{4Dc^{2m}t'}\right).
\end{eqnarray}
Here we used the abbreviation
\begin{equation}
\label{eqn:C_definition}
\mathcal{C}_{n,m}^{(\alpha)}=\frac{1}{\left(c^{-\alpha};c^{-\alpha}\right)_m
\left(c^\alpha;c^\alpha\right)_{n-m}}.
\end{equation}
The integral over time can be performed analytically, yielding
\begin{eqnarray}
\nonumber
\fl p_r(x,t|x_0)&=&e^{-rt}\frac{1}{\sqrt{4\pi Dt}}\exp\left(-\frac{(x-x_0)^2}
{4Dt}\right)+re^{-rt}\sum_{n=1}^{\infty}\frac{(rt)^{n-1}}{2 D(n-1)!\Gamma\left(
n+\frac{1}{2}\right)}\\
\nonumber
\fl&&\times\sum_{m=0}^n\mathcal{C}^{(2)}_{n,m}c^{-2m}\biggr[(n-1)!c^m\sqrt{Dt}\,
_1F_1\left(-n+\frac{1}{2};\frac{1}{2};-\frac{\left(x-c^nx_0\right)^2}{4Dc^{2m} t}\right)+\\
\label{eqn:Brownian_explicit}
\fl&&-\Gamma\left(n+\frac{1}{2}\right)\left|x-c^nx_0\right|\,_1F_1\left(1-n;
\frac{3}{2};-\frac{\left(x-c^nx_0\right)^2}{4Dc^{2m}t}\right)\biggr],
\end{eqnarray}
where $_1F_1$ denotes the Kummer confluent hypergeometric function. We note that
in \cite{shlomi}, the authors derived the Fourier-Laplace transform of the
propagator for the special initial condition $x_0=0$. Our results above
extend this result to an arbitrary initial condition and we invert this general
form to real space.

\begin{figure}
\centering
(a)\includegraphics[scale=0.30]{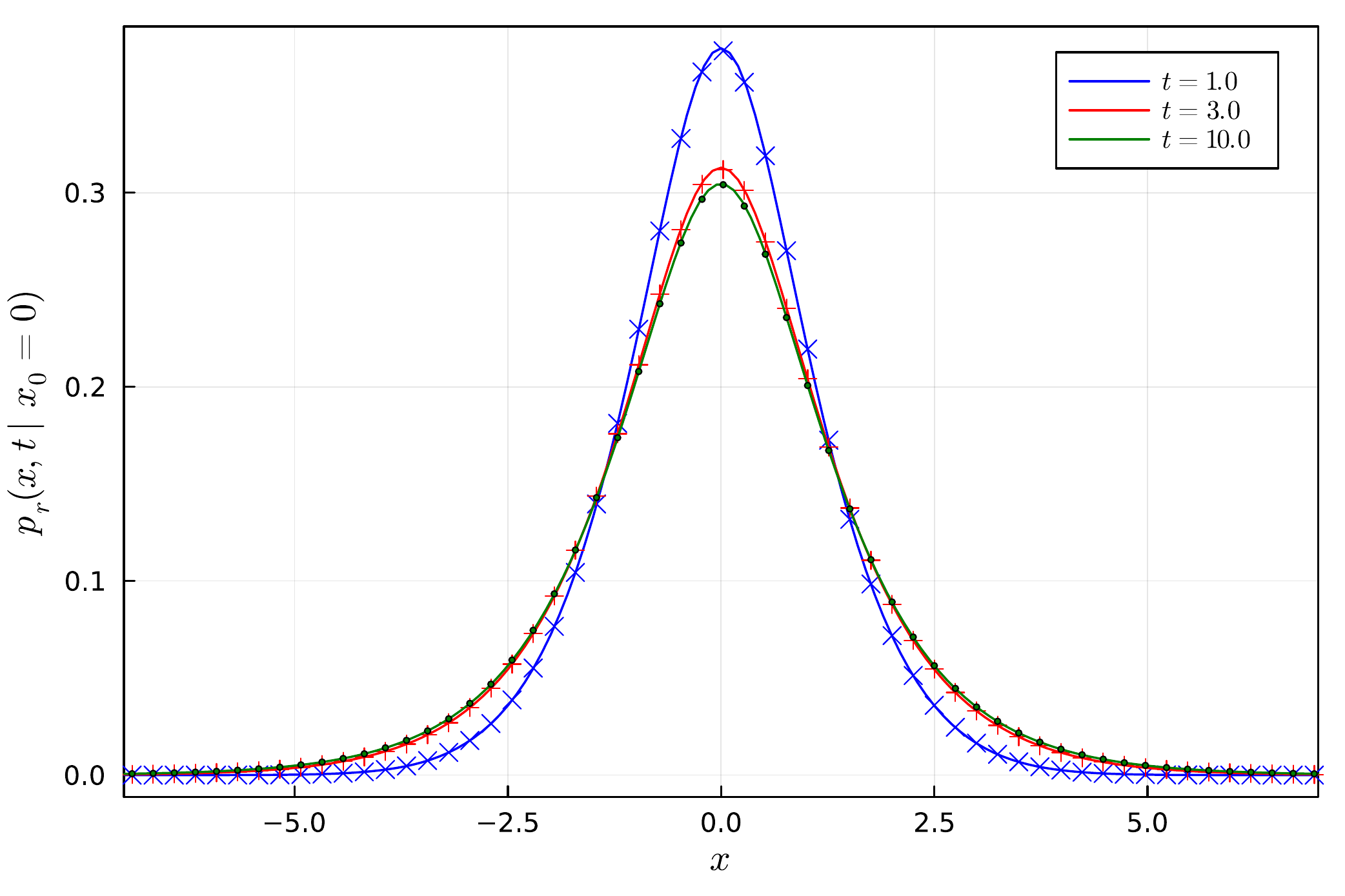}
(b)\includegraphics[scale=0.30]{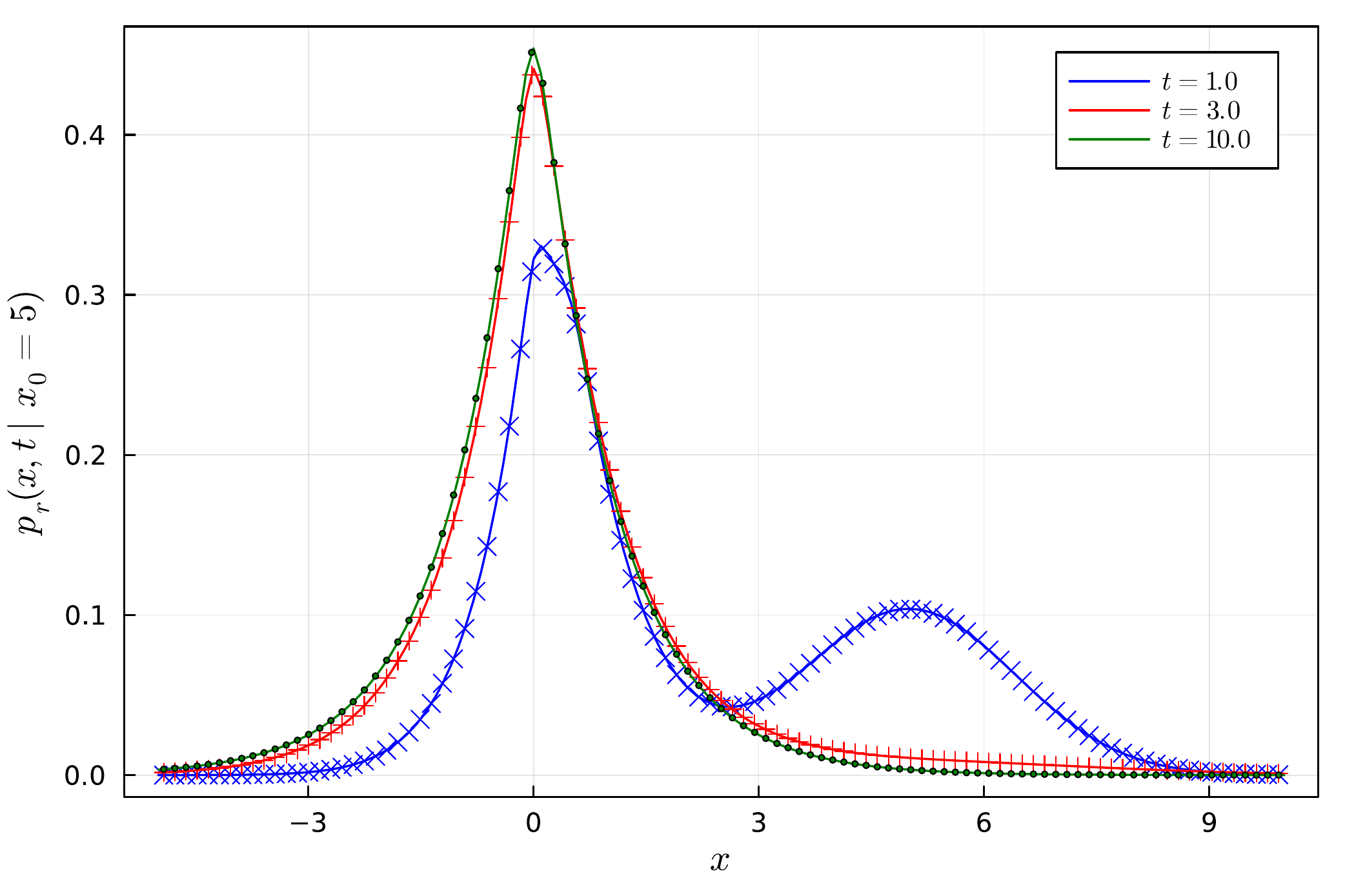}
(c)\includegraphics[scale=0.30]{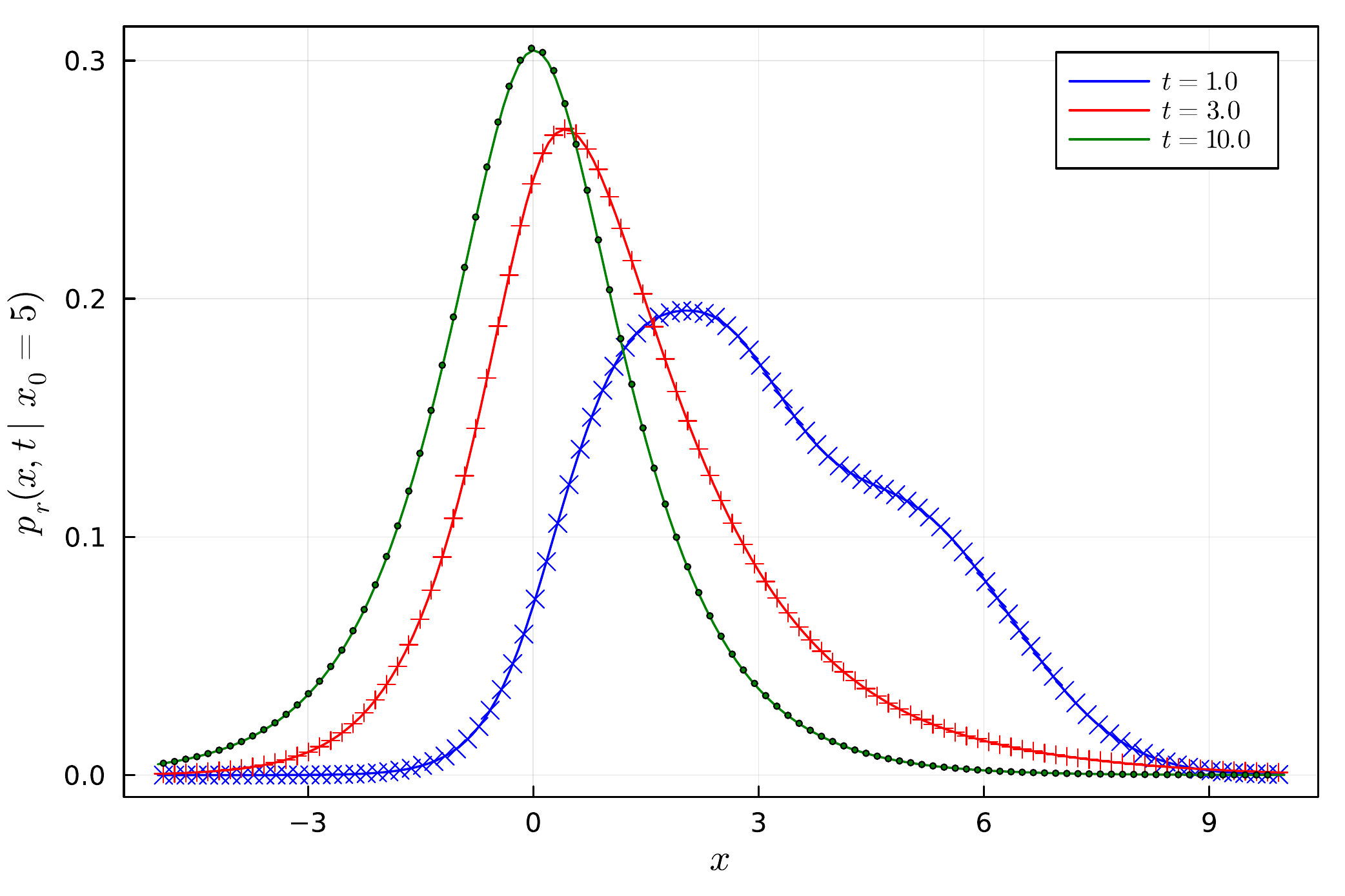}
(d)\includegraphics[scale=0.30]{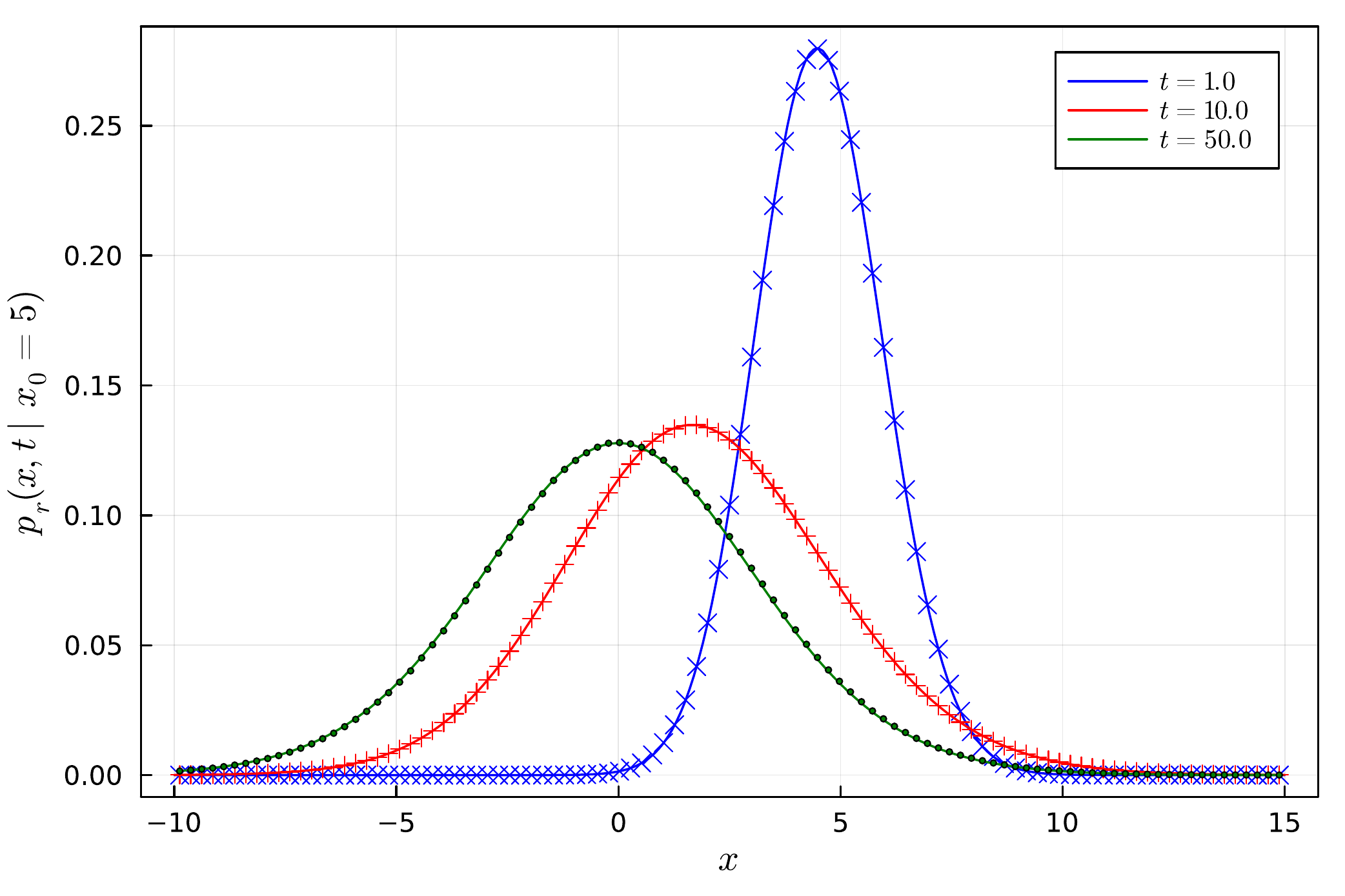}
\caption{Propagator for Brownian motion with partial resetting. Solid lines
represent the analytical distribution (\ref{eqn:Brownian_explicit}), while
symbols represent the simulations results. In (a) the starting position is
$x_0=0$, while in (b)-(d) it is $x_0=5$. The resetting factor $c$ was chosen
as $c=0.5$ in (a), $c=0.1$ in (b), $c=0.5$ in (c), and $c=0.9$ in (d).}
with time.
\label{fig:Brownian_PSR1}
\end{figure}

\begin{figure}
\centering
\includegraphics[scale=0.30]{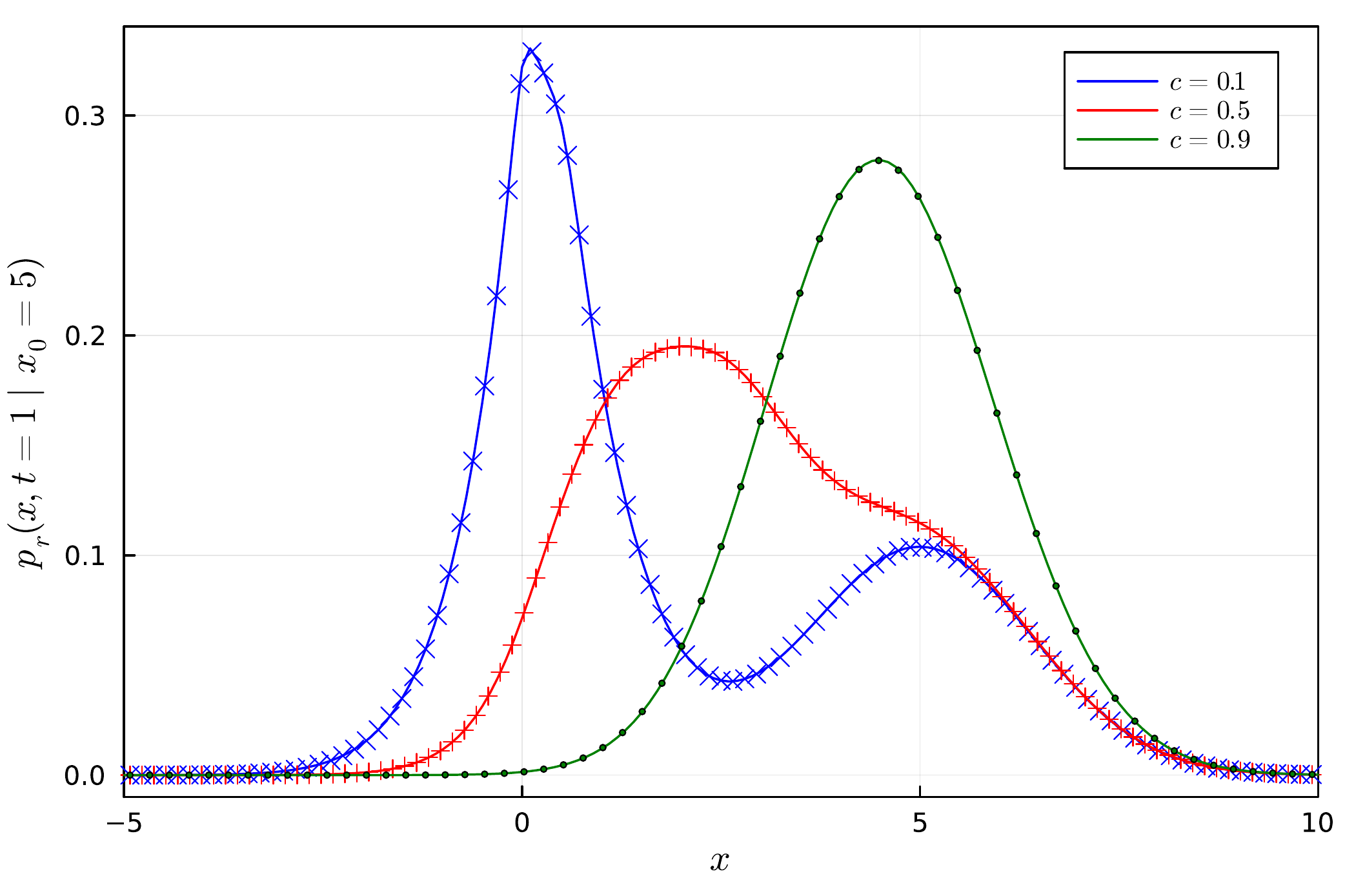}
\includegraphics[scale=0.30]{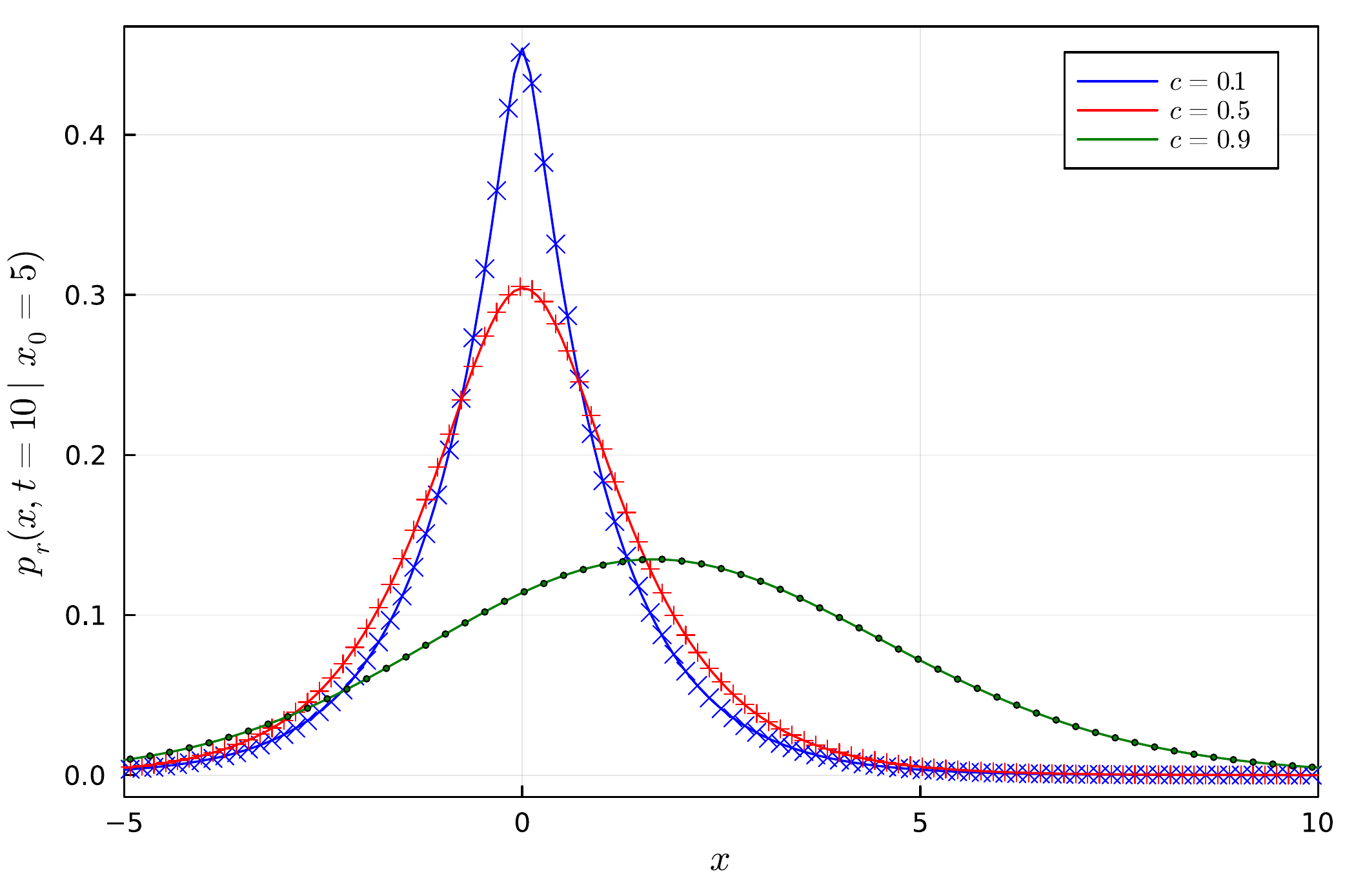}
\caption{Propagator for Brownian motion with PSR with starting position $x_0=5$,
and for different values of the resetting factor $c$. The time is chosen as (a)
$t=1$ and (b) $t=10$.}
\label{fig:Brownian_PSR2}
\end{figure}

The PDF (\ref{eqn:Brownian_explicit}) is shown in figures \ref{fig:Brownian_PSR1}
and \ref{fig:Brownian_PSR2} for different choices of the parameters. The agreement
between theory and simulations is excellent. The simulated PDF was obtained with
the algorithm described in \ref{appendix:numerical}. We note that while the PDF
stays symmetric around the origin when the process is initiated in $x=0$, when
the initial condition is away from the origin, strong asymmetries of the PDF are
observed. These asymmetries relax as function of time, and eventually convergence
to a stationary, symmetric form. As seen in figure \ref{fig:Brownian_PSR1} the
relaxation to stationarity requires more time when the resetting factor $c$ is
closer to the value $c=1$ in absence of any resetting.

To obtain a concrete form for the stationary distribution, we set $\alpha=2$ in
equation \eqref{eqn:stat_Fox} to get
\begin{equation}
p^{(s)}_r(x)=\frac{1}{2(c^2;c^2)_{\infty}}\sqrt{\frac{r}{D}}\sum_{n=0}^{\infty}
\frac{c^{-n}}{(c^{-2};c^{-2})_{n}} e^{-\sqrt{r/D}\ c^{-n}|x|},
\end{equation}
which is in agreement with \cite{pierce}.

\section{PSR for the Cauchy case}

In the case $\alpha=1$, for which the L{\'e}vy stable density is given by the
Cauchy (Lorentz) distribution, the function $p_0$ reads
\begin{equation}
p_0(x,t)=\dfrac{1}{\pi}\dfrac{Dt}{x^2 + D^2t^2}.
\end{equation}
With equation \eqref{eqn:pr_Levy_flight} we therefore find
\begin{equation}
p_r(x,t|x_0)=\dfrac{Dt}{x^2+D^2t^2}+re^{-rt}\sum_{n=1}^{\infty}\int_0^tdt'
\dfrac{\left[r(t-t')^{n-1}\right]}{\pi(n-1)!}\sum_{m=0}^n\mathcal{C}_{n,
m}^{(1)}\dfrac{1}{\pi}\dfrac{Dc^mt'}{(x-c^nx_0)^2+(Dc^mt')^2}.
\end{equation}
The integral can be performed analytically, yielding
\begin{eqnarray}
\nonumber
p_r(x,t| x_0)&=&\dfrac{1}{\pi}\dfrac{Dt}{x^2+D^2t^2}+\dfrac{1}{\pi}e^{-rt}
\sum_{n=1}^{\infty} \dfrac{(rt)^n}{(n+1)!}\sum_{m=0}^n\mathcal{C}_{n,m}^{(1)}
\dfrac{D c^m t}{(x-c^n x_0)^2}\\
&&\times\,_3F_2\left(1,1,\frac{3}{2};\frac{n}{2}+1,\frac{n}{2}+\frac{3}
{2};\dfrac{-D^2 c^{2 m} t^2}{(x-c^n x_0)^2}\right),
\label{eqn:Cauchy_explicit}
\end{eqnarray}
where $_pF_q(a_1,\dots,a_p;b_1,\dots,b_p;z)$ is the generalised hypergeometric
function \cite{special_functions}. The stationary PDF follows from equation
\eqref{eqn:stat_distrib_general}\footnote{We may alternatively set $\alpha=1$
in equation \eqref{eqn:stat_Fox}}. The integral for $\alpha=1$ is computed
explicitly in reference \cite{kusmierz}, and we find
\begin{eqnarray}
\nonumber
p^{(s)}_r(x)&=&\dfrac{1}{\pi(c^\alpha;c^\alpha)_{\infty}}\sum_{n=0}^{\infty}
\dfrac{1}{(c^{-\alpha};c^{-\alpha})_n}\dfrac{\lambda}{c^n}\\
&&\times\left[\left(\dfrac{\pi}{2}-\text{Si}\left(\dfrac{\lambda|x|}{c^n}\right)
\right)\sin\left(\dfrac{\lambda|x|}{c^n}\right)-\cos\left(\dfrac{\lambda x}{
c^n}\right)\text{Ci}\left(\dfrac{\lambda|x|}{c^n}\right)\right],
\label{eqn:cauchy_stat}
\end{eqnarray}
where we again used $\lambda=r/D$, and where the sine/cosine integrals $\text{
Si}(x)$ and $\text{Ci}(x)$ are defined as
\begin{equation}
\text{Si}(x)=\int_0^x\dfrac{\sin(t)}{t}dt,\quad\text{Ci}(x)=\int_x^{\infty}
\dfrac{\cos(t)}{t}dt.
\end{equation}
In the case of total resetting with $c=0$, this result coincides with the one
obtained in \cite{kusmierz}.

\begin{figure}
\centering
(a)\includegraphics[width=7.2cm]{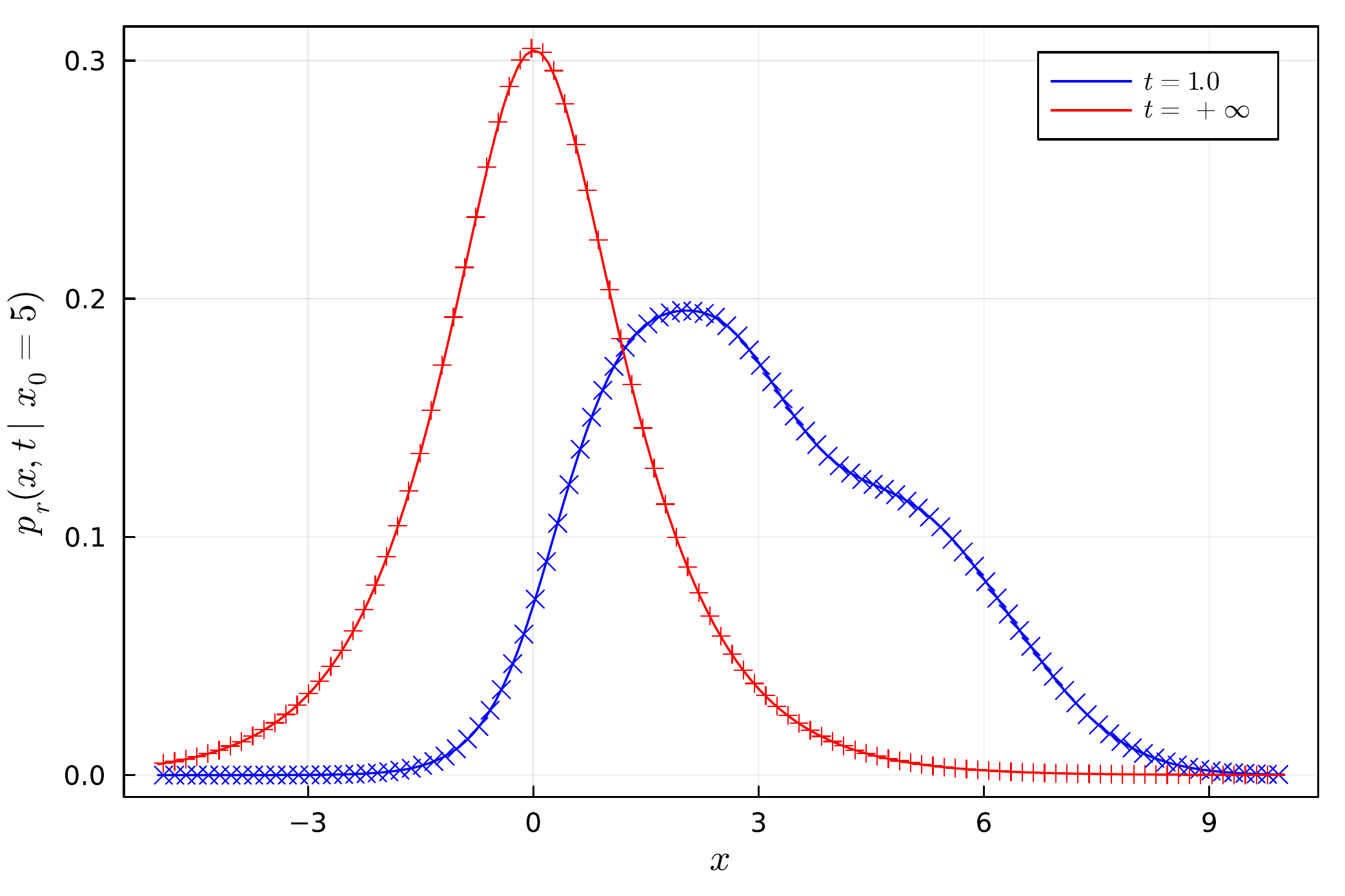}
(b)\includegraphics[width=7.2cm]{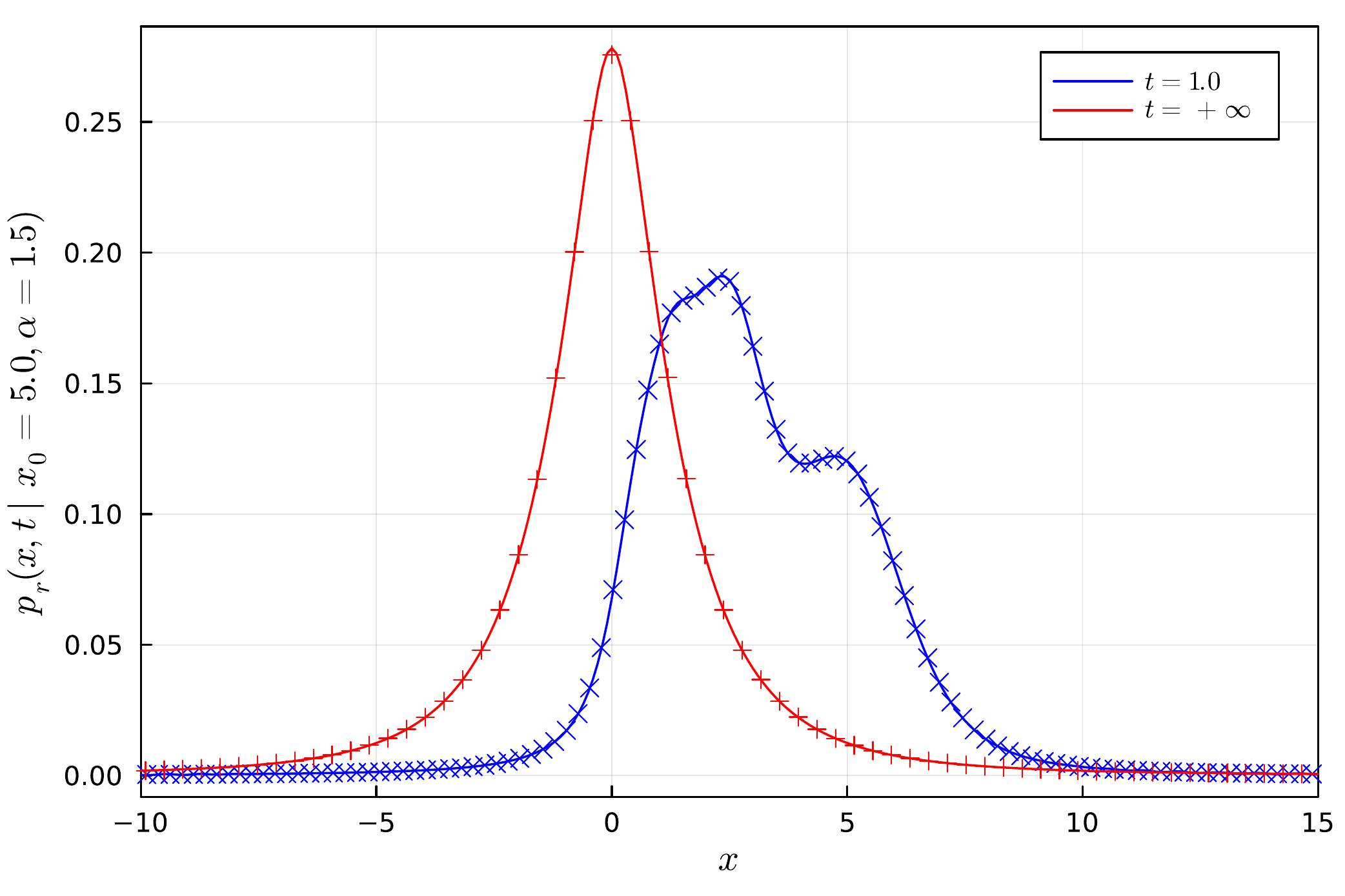}
(c)\includegraphics[width=7.2cm]{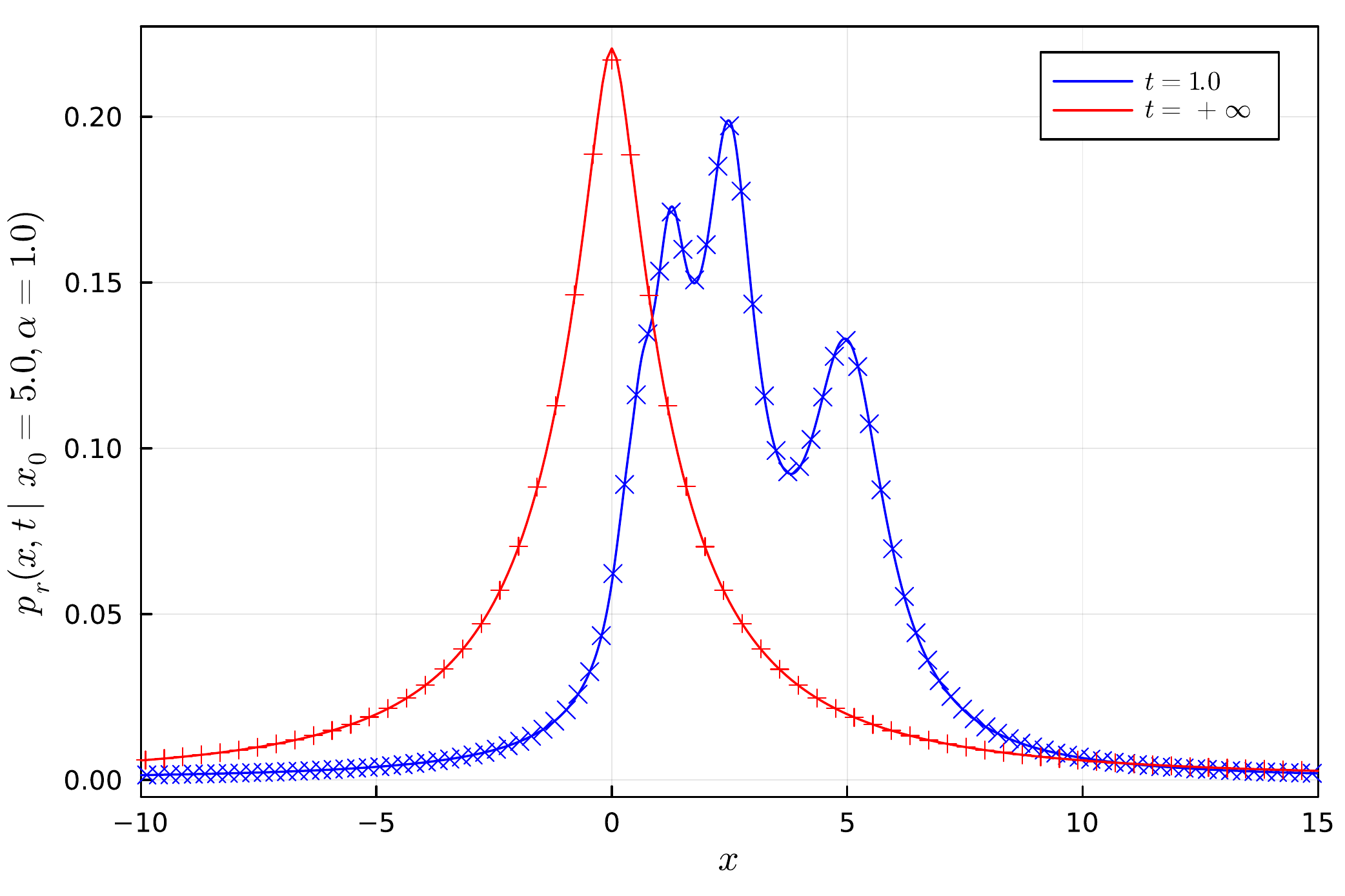}
(d)\includegraphics[width=7.2cm]{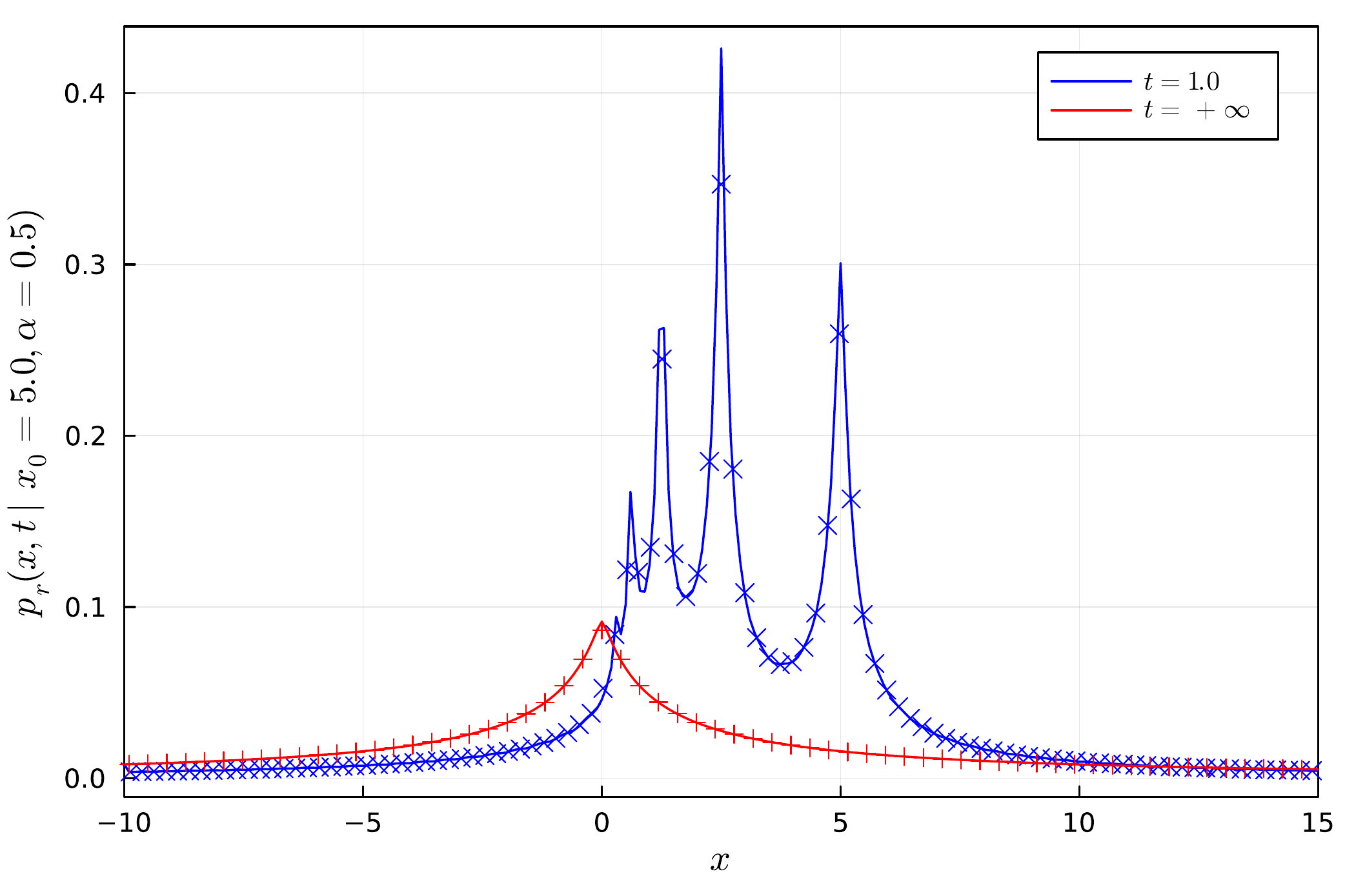}
\caption{Propagator for Brownian motion and L{\'e}vy flight with PSR for (a)
$\alpha=2$ (Brown), (b) $\alpha=1.5$, (c) $\alpha=1.0$ (Cauchy), and (d) $\alpha
=0.5$. Parameters: $x_0=5$, $D=1$, $r=1$, $t=1$, and $c=0.5$.}
\label{fig:Cauchy_PSR}
\end{figure}

In figure \ref{fig:Cauchy_PSR} we plot the PDF \eqref{eqn:Cauchy_explicit} and
the associated stationary PDF \eqref{eqn:cauchy_stat} along with examples for
Gaussian and other L{\'e}vy flight processes with $\alpha=2.0$, $\alpha=1.5$,
and $\alpha=0.5$. For the latter cases expression (\ref{eqn:pr_Levy_flight})
was evaluated numerically. The agreement with numerical simulations is
excellent in all cases. A distinct feature are the strong asymmetries in the
PDF due to the initial condition. For lower $\alpha$, i.e., longer-tailed
stable densities, the multimodal structure becomes more spiky. The multimodality
of the PDF can be anticipated from equation \eqref{eqn:series_expansion_pr},
according to which the PDF is a sum of single-peaked functions centred on
different positions.

\section{Conclusions}

We established a framework to calculate the time-dependent PDF in the presence
of partial resetting effects for homogeneous Markov processes with Poissonian
resetting, in which the process is partially reset by multiplication with
a constant factor $0<c<1$ at random times. We showed that, consistently,
the limiting cases $c\to0$ and $c\to1$ of this model correspond to total
resetting \cite{masaprl} and a stochastic process without resetting,
respectively. We derived an exact representation of the PDF in the real
space-time domain for the case of general symmetric L{\'e}vy flights with
stable index $0<\alpha\le2$, including Brownian motion and Cauchy flights as
particular cases for $\alpha=2$ and $1$. As our approach is valid for generic
Markov processes, in the future other densities such as asymmetric L{\'e}vy
stable forms can be studied. For the case of non-zero initial conditions, we
reported highly asymmetric non-stationary PDFs for $\alpha=2$ and the emergence
of non-trivial inhomogeneous multimodal regimes with $\alpha\neq2$. We also
determined the stationary PDF for symmetric L{\'e}vy flights in terms of
Fox $H$-functions and presented the particular cases $\alpha=2$ and $1$ in
terms of elementary functions. Moreover, we also showed how the resetting
factor $c$ influences the relaxation speed towards stationarity.

We expect that our results will find applications in systems ranging from the
generic theory of search processes over financial mathematics to population
dynamics and geophysics.  In the future it will be relevant to work out the
precise relaxation dynamics towards the steady state and the tails of the
PDFs under PSR dynamics. Moreover, it will be important to determine the
associated first-passage behaviour. Finally, as another challenge we mention
the description of non-Markov PSR-processes.

\appendix

\section{Solution of last renewal equation}
\label{appendix:last_renewal}

As stated in the main text, the solution of the last renewal equation
\eqref{eqn:last_renewal_equation} is given by expressions
\eqref{eqn:series_expansion_pr} and \eqref{eqn:qn_definition}. We now check
this by inserting \eqref{eqn:series_expansion_pr} and \eqref{eqn:qn_definition}
into \eqref{eqn:last_renewal_equation},
\begin{eqnarray}
\nonumber
&&e^{-rt}\sum_{n=0}^{\infty}r^nq_n(x,t|x_0)=\\
\nonumber
&&\qquad=e^{-rt}q_0(x,t|x_0)+r\int_0^t\mathrm{d}t'e^{-rt'}\int_{-\infty}^{\infty}
\mathrm{d}ye^{-r(t-t')}\sum_{n=0}^{\infty}r^nq_n(y,t-t'|x_0)p_0(x,t'|cy)\\
\nonumber
&&\qquad=e^{-rt}q_0(x,t|x_0)+re^{-rt}\sum_{n=0}^{\infty}r^n\int_0^t\mathrm{d}t'
\int_{-\infty}^{\infty}\mathrm{d}yq_n(y,t-t'|x_0)p_0(x,t'|cy)\\
\nonumber
&&\qquad=e^{-rt}q_0(x,t|x_0)+e^{-rt}\sum_{n=0}^{\infty}r^{n+1}q_{n+1}(x,t|x_0)\\
&&\qquad=e^{-rt}\sum_{n=0}^{\infty}r^nq_n(x,t|x_0)
\end{eqnarray}
which completes the proof.

\section{Fourier-Laplace identity}
\label{appendix:Fourier_Laplace}

At the end of section \ref{section:general_derivation}, we used the following
identity when we were checking the limit $c\to1$,
\begin{equation}
\label{eqn:Fourier_Laplace_identity}
\dfrac{\hat{\tilde{p}}_0(k,r+s)}{1-r\hat{\tilde{p}}_0(k,r+s)}=\hat{\tilde{p}}
_0(k,s),
\end{equation}
which is valid for time and space-homogeneous propagators. First we point out
that the right hand side does not depend on $r$. This should not be surprising:
we are considering the limit in which PSR does not affect the motion, hence the
rate $r$ should not play any role in this case. Nevertheless, this identity is
indeed valid for general L{\'e}vy processes. It can be proved by using the
L{\'e}vy-Khinchine theorem \cite{sato} which gives an analytical general
expression for the characteristic function of L{\'e}vy process,
\begin{equation}
\hat{p}_0(k,t)=\exp\left(t\left(aik-\frac{1}{2}\sigma^2k^2+\int_{\mathbb{R}
\setminus\{0\}}\left(e^{ikx}-1-ikx\mathbf{1}_{|x|<1}\right)\Pi(d x)\right)\right),
\end{equation}
where $a\in\mathbb{R}$, $\sigma\geq0$, and $\Pi$ is the L{\'e}vy measure of the
process. Hence, the Laplace transform of this expression has the following form
\begin{equation}
\hat{\tilde{p}}_0(k,s)=\dfrac{1}{s+f(k)},
\end{equation}
for some function $f(k)$. Let us substitute this expression into the left hand
side of \eqref{eqn:Fourier_Laplace_identity},
\begin{equation}
\dfrac{1}{r+s+f(k)}\left(1-\dfrac{r}{r+s+f(k)}\right)^{-1}=\dfrac{1}{s+f(k)},
\end{equation}
so that we showed that the left and right hand sides are identical.

\section{Equivalence between first renewal and Fokker-Planck equation}
\label{appendix:equivalence}

We stated in the main text that the system may be equivalently described via the
fractional Fokker-Planck equation (FPE) \eqref{eqn:Fokker-Planck_equation}. We
show here that the solution \eqref{eqn:hat_tilde_pr_Levy_flights} we obtained
for L{\'e}vy flights is indeed a solution of the FPE. The FPE in Fourier-Laplace
space reads
\begin{equation}
-e^{ikx_0}+s\hat{\tilde{p}}_r(k,s|x_0)=-(r+D|k|^\alpha)\hat{\tilde{p}}_r(k,s|
x_0)+r\hat{\tilde{p}}_r(kc,s|x_0),
\end{equation}
which can be rearranged in the form
\begin{equation}
\label{eqn:alternative}
(r+s+D|k|^\alpha)\hat{\tilde{p}}_r(k,s|x_0)=e^{ikx_0}+r\hat{\tilde{p}}_r(kc,s|x_0).
\end{equation}
Substituting \eqref{eqn:hat_tilde_pr_Levy_flights} in the last equation we get
\begin{equation}
\sum_{n=0}^{\infty}r^ne^{ikc^nx_0}\prod_{l=1}^n\dfrac{1}{r+s+Dc^{\alpha l}|k|^
\alpha}=e^{ikx_0}+r\sum_{n=1}^{\infty}r^ne^{ikc^nx_0}\prod_{l=1}^n\dfrac{1}{r+s
+Dc^{\alpha l}|k|^\alpha},
\end{equation}
where we changed the summation index and use the convention that the empty
product $\prod_{l=1}^0=1$. Alternatively, we could have derived equation
\eqref{eqn:hat_tilde_pr_Levy_flights} from \eqref{eqn:alternative}. Nevertheless,
we preferred adopting the more general equation
\eqref{eqn:propagator_general_result} for the specific case of symmetric
L{\'e}vy flights.

\section{Normalisation identity}
\label{appendix:identity}

When we discussed normalisation we encountered the identity \eqref{identity}.
This identity is an immediate consequence of the $q$-binomial theorem and may
be proved by using corollary (c) in section 10.2.2 of \cite{special_functions}.
Indeed, we know from this reference that the following result holds for $|x|<1$
and $|q|<1$,
\begin{equation}
\sum_{m=0}^n\dfrac{(q;q)_n}{(q;q)_m(q;q)_{n-m}}(-1)^mq^{\frac{1}{2}m(m-1)}x^m=
(x;q)_n.
\end{equation}
If we set $x=q$ in the previous formula we get
\begin{equation}
\sum_{m=0}^n\dfrac{(q;q)_n}{(q;q)_m(q;q)_{n-m}}(-1)^mq^{\frac{1}{2}m(m+1)}=(q;q)_n.
\end{equation}
The factor $(q;q)_n$ can now be simplified on both sides, and the general term
of the summation can be rewritten as
\begin{equation}
\sum_{m=0}^n\dfrac{1}{(q;q)_m(q;q)_{n-m}}(-1)^mq^{\frac{1}{2}m(m+1)}=\sum_{m=0}^n
\dfrac{1}{(q;q)_{n-m}}\left(\prod_{l=1}^m\dfrac{1}{1-q^l}\right)\dfrac{(-1)^m}{q^
{-\frac{1}{2}m(m+1)}}=1,
\end{equation}
where we used the definition of the $q$-Pochhammer symbol to rewrite the term
$(q;q)_m$. We now notice that
\begin{equation}
\left(\prod_{l=1}^m\dfrac{1}{1-q^l}\right)\dfrac{(-1)^m}{q^{-\frac{1}{2}m(m+1)}}
=\prod_{l=1}^m\dfrac{1}{1-q^{-l}}=\dfrac{1}{(q^{-1};q^{-1})_m}.
\end{equation}
Therefore,
\begin{equation}
\sum_{m=0}^n\dfrac{1}{(q^{-1};q^{-1})_m(q;q)_{n-m}}=1,
\end{equation}
which completes the proof.

\section{Numerical simulations}
\label{appendix:numerical}

To confirm our analytical results, we simulated the stochastic process and
compared the results with the analytical distributions given in equations
\eqref{eqn:pr_Levy_flight} and \eqref{eqn:stat_distrib_general}. Numerical
simulations are pretty straightforward and simple. Nevertheless, for the sake
of clarity, we include a brief summary of the simulations strategy. The
algorithm to generate the random variable $Y_t$ is
\begin{enumerate}
\item sample the total number of partial resetting events $N_t$ from a
Poissonian distribution with rate $r$;
\item sample $T_1,T_2,\dots,T_{N_t}$ uniformly on the interval $[0,t]$. This
is equivalent to sampling all random variables $\lbrace T_i\rbrace_{i=1}^{N_t}$
until saturation of the total time $t$;
\item sample displacements between partial resetting events $X_{T_1},\dots,
X_{T_{N_t}}$ from a symmetric $\alpha$-stable distribution by using the
library \texttt{AlphaStableDistributions.jl} available in the Julia language
package;
\item use equation \eqref{psr} to compute $Y_t$.
\end{enumerate}
In all cases we sampled $5\cdot10^7$ values of the random variable $Y_t$ and
we computed a histogram. Therefore, our naive algorithm only allows the
sampling of typical values of the random variable $Y_t$.

Concerning the analytical distribution, we directly implemented equations
\eqref{eqn:pr_Levy_flight} and \eqref{eqn:stat_distrib_general} computing
the integral with adaptive Gauss-Kronrod integration as implemented in
the GNU Scientific Library (GSL). Since the integrand value of $p_0$
is often very small, we instead calculated the integral for
$\exp(\kappa)p_0(x-c^nx_0,c^{\alpha m}t)$ with suitable shift, e.g.,
$\kappa=(x-c^nx_0)^2/(4c^{\alpha m}t)$ for a Gaussian, and multiplied the
integral with $\exp(-\kappa)$ afterwards. For this purpose and for performing
the sums, also because the summation terms often have alternating sign and
strongly varying magnitudes, we used the high precision library {\tt
mpfr} with 200 bits precision for these operations. We also point
out that, concerning the Brownian and the Cauchy cases, the explicit formulas
\eqref{eqn:Brownian_explicit} \eqref{eqn:Cauchy_explicit} are not easy to
compute due to the poor implementation of hypergeometric functions. To the
authors' knowledge, this issue is common in many programming languages.

\ack

We acknowledge funding from the German Science Foundation (DFG, grant
no. ME 1535/12-1). AVC acknowledges the support of the Polish National
Agency for Academic Exchange (NAWA). Z. Palmowski acknowledges that the
research is partially supported by Polish National Science Centre Grant
No. 2021/41/B/HS4/00599.


\section*{References}

\begin{thebibliography}{99}

\bibitem{landau} E. M. Lifshitz and L. P. Pitaevski, Landau and Lifshitz
Course of Theoretical Physics 10: Physical Kinetics (Butterworth-Heinemann,
Oxford UK, 1981).

\bibitem{vankampen} N. van Kampen, Stochastic processes in physics and chemistry
(North Holland, Amsterdam, 1981).

\bibitem{levy} P. L{\'e}vy, Processus stochastiques et mouvement brownien
(Gauthier-Villars, Paris, 1948).

\bibitem{brenig} W. Brenig, Statistical theory of heat: Nonequilibrium
phenomena (Springer, Berlin, 1989).

\bibitem{haenggirev} J. Spiechowicz, I. G. Marchenko, P. H{\"a}nggi, and J.
{\L}uczka, Diffusion coefficient of a Brownian particle in equilibrium and
nonequilibrium: Einstein model and beyond, Entropy \textbf{25}, 42 (2023).

\bibitem{franosch} F. H\"ofling and T. Franosch, Anomalous transport in the
crowded world of biological cells, Rep. Prog. Phys. \textbf{76}, 046602 (2013).

\bibitem{vilk1} O. Vilk, E. Aghion, T. Avgar, C. Beta, O. Nagel, A. Sabri, R.
Sarfati, D. K. Schwartz, M. Weiss, D. Krapf, R. Nathan, R. Metzler, and M.
Assaf, Unravelling the origins of anomalous diffusion: from molecules to
migrating storks, Phys. Rev. Res. \textbf{4}, 033055 (2022).

\bibitem{loewen} C. Bechinger, R. Di Leonardo, H. L{\"o}wen, C. Reichhardt,
G. Volpe, and G. Volpe, Active particles in complex and crowded environments,
Rev. Mod. Phys. \textbf{88}, 045006 (2016).

\bibitem{brianrev} B. Berkowitz, Characterizing flow and transport in fractured
geological media: a review, Adv. Wat. Res. \textbf{25}, 861 (2002).

\bibitem{scher} H. Scher and E. W. Montroll, Anomalous transit-time dispersion
in amorphous solids, Phys. Rev. B \textbf{12}, 2455 (1975).

\bibitem{bouchaud} J.-P. Bouchaud and M. Potters, Theory of financial risk
and derivative pricing: From statistical physics to risk management
(Cambridge University Press, Cambridge UK, 2000).

\bibitem{brockmann} D. Brockmann and D. Helbing, The hidden geometry of complex,
network-driven contagion phenomena, Science \textbf{342}, 1337 (2013).

\bibitem{benichourev} O. B{\'e}nichou, C. Loverdo, M. Moreau, and R. Voituriez,
Intermittent search strategies, Rev. Mod. Phys. \textbf{83}, 81 (2011).

\bibitem{carlos} T. Mattos, C. Mej\'{\i}a-Monasterio, R. Metzler, and G.
Oshanin, First passages in bounded domains: When is the mean first passage
time meaningful?, Phys. Rev. E \textbf{86}, 031143 (2012).

\bibitem{aljaz} A. Godec and R. Metzler, Universal proximity effect in target
search kinetics in the few encounter limit, Phys. Rev. X \textbf{6}, 041037
(2016).

\bibitem{denis} D. Grebenkov, R. Metzler, and G. Oshanin, Strong defocusing of
molecular reaction times: geometry and reaction control, Commun. Chem.
\textbf{1}, 96 (2018).

\bibitem{bvh} P. H. von Hippel, and O. Berg, Facilitated target location in
biological systems, J. Biol. Chem. \textbf{264}, 675 (1989).

\bibitem{gijs} M. A. Lomholt, B. v. d. Broek, S.-M. J. Kalisch, G. J. L. Wuite,
and R. Metzler, Facilitated diffusion with DNA coiling, Proc. Natl. Acad. Sci.
USA \textbf{106}, 8204 (2009).

\bibitem{adam} G. Adam and M. Delbr{\"u}ck, Reduction of dimensionality in
biological diffusion processes, in Structural Chemistry and Molecular
Biology, edited by A. Rich and N. Davidson (Freeman, San Francisco, CA, 1968).

\bibitem{mirnyjpa} L. Mirny, M. Slutsky, Z. Wunderlich, A. Tafvizi, J. Leith,
and A. Kosmrlj, How a protein searches for its site on DNA: the mechanism of
facilitated diffusion, J. Phys. A \textbf{42}, 434013 (2009).

\bibitem{ghandibook} G. E. Viswanathan, M. G. E. da Luz, E. P. Raposo, and H. E.
Stanley, The physics of foraging: an introduction to random searches and
biological encounters (Cambridge University Press, Cambridge UK, 2011).

\bibitem{sims} D. W. Sims, et al., Scaling laws of marine predator search
behaviour, Nature \textbf{451}, 1098 (2008).

\bibitem{vladpnas} V. V. Palyulin, A. V. Chechkin, and R. Metzler, L{\'e}vy
flights do not always optimize random blind search for sparse targets,
Proc. Natl. Acad. Sci. USA \textbf{111}, 2931 (2014).

\bibitem{vladjstat} V. V Palyulin, A. Chechkin, and R. Metzler, Space-fractional
Fokker-Planck equation and optimization of random search processes in the
presence of an external bias, J. Stat. Mech. \textbf{2014}, P11031 (2014).

\bibitem{vladjpa} V. V. Palyulin, A. Chechkin, R. Klages, and R. Metzler, Search
reliability and search efficiency of combined L{\'e}vy-Brownian motion: long
relocations mingled with thorough local exploration, J. Phys. A \textbf{49},
394002 (2016).

\bibitem{dybiecjpa} B. Dybiec, E. Gudowska-Nowak, and A. Chechkin, To hit or
to pass it over---remarkable transient behavior of first arrivals and passages
for L{\'e}vy flights in finite domains, J. Phys. A \textbf{49}, 504001 (2016).

\bibitem{vladepjb} V. V. Palyulin, V. N. Mantsevich, R. Klages, R. Metzler, and
A. Chechkin, Comparison of pure and combined search strategies for single and
multiple targets, Eur. Phys. J. B \textbf{90}, 170 (2017).

\bibitem{vladnjp} V. V. Palyulin, G. Blackburn, M. A. Lomholt, N. W. Watkins,
R. Metzler, R. Klages, and A. Chechkin,  First-passage and first-hitting times
of L{\'e}vy flights and L{\'e}vy walks, New J. Phys. \textbf{21} 103028 (2019).

\bibitem{amin} A. Padash, T. Sandev, H. Kantz, R. Metzler, and A. Chechkin,
Asymmetric L{\'e}vy Flights Are More Efficient in Random Search, Fractal Fract
\textbf{6}, 260 (2022).

\bibitem{mich} M. A. Lomholt, T. Ambj{\"o}rnsson, and R. Metzler,
Optimal target search on a fast folding polymer chain with volume exchange,
Phys. Rev. Lett. \textbf{95}, 260603 (2005).

\bibitem{olivier} O. B{\'e}nichou, M. Coppey, M. Moreau, P. H. Suet, and R.
Voituriez, Phys. Rev. Lett. \textbf{94}, 198101 (2005).

\bibitem{olivier1} O. B{\'e}nichou, C. Loverdo, M. Moreau, and R. Voituriez,
Phys. Rev. E \textbf{74}, 020102 (2006).

\bibitem{heiko} K. Schwarz, Y. Schr{\"o}der, and H. Rieger, Phys. Rev. E
\textbf{94}, 042133 (2016).

\bibitem{masaprl} M. R. Evans and S. N. Majumdar, Diffusion with stochastic
resetting, Phys. Rev. Lett. \textbf{106}, 160601 (2011).

\bibitem{masajpa} M. R. Evans and S. N. Majumdar, Diffusion with Optimal
Resetting, J. Phys. A \textbf{44}, 435001 (2011).

\bibitem{besga} B. Besga, A. Bovon, A. Petrosyan, S. N. Majumdar, and S.
Ciliberto, Optimal mean first-passage time for a Brownian searcher subjected
to resetting: Experimental and theoretical results, Phys. Rev. Res. \textbf{2},
032029(R) (2020).

\bibitem{pal} A. Pal, A. Kundu, and M. R. Evans, Diffusion under time-dependent
resetting, J. Phys. A \textbf{49}, 225001 (2016).

\bibitem{bhat} U. Bhat, C. De Bacco, and S. Redner, Stochastic search with
Poisson and deterministic resetting, J. Stat. Mech. \textbf{2016}, 083401 (2016).

\bibitem{sr1} T. Rotbart, S. Reuveni, and M. Urbakh, Michaelis-Menten reaction
scheme as a unified approach towards the optimal restart problem, Phys. Rev. E
\textbf{92}, 060101(R) (2015).

\bibitem{sr2} S. Reuveni, Optimal stochastic restart renders fluctuations in
first passage times universal, Phys. Rev. Lett. \textbf{116}, 170601 (2016).

\bibitem{max} C. Godr{\`e}che and J.-M. Luck, Maximum and records of random
walks with stochastic resetting, J. Stat. Mech. \textbf{2022}, 063202 (2022).

\bibitem{cheso} A. V. Chechkin and I. M. Sokolov, Random search with resetting:
A unified renewal approach, Phys. Rev. Lett. \textbf{121}, 050601 (2018).

\bibitem{igorlr} I. M. Sokolov, Linear response and fluctuation-dissipation
relations for Brownian motion under resetting, Phys. Rev. Lett. \textbf{130},
067101 (2023).

\bibitem{quant} S. Wald and L. B{\"o}ttcher, From classical to quantum walks
with stochastic resetting on networks, Phys. Rev. E \textbf{103}, 012122 (2021).

\bibitem{martinrev} M. R. Evans, S. N. Majumdar, and G. Schehr,
Stochastic resetting and applications, J. Phys. A \textbf{53}, 193001 (2020).

\bibitem{exp} O. Tal-Friedman, A. Pal, A. Sekhon, S. Reuveni, and Y. Roichman,
Experimental Realization of Diffusion with Stochastic Resetting, J. Phys. Chem.
Lett. \textbf{11}, 7350 (2020).

\bibitem{exp1} B. Besga, A. Bovon, A. Petrosyan, S. N. Majumdar, and S.
Ciliberto, Optimal mean first-passage time for a Brownian searcher subjected
to resetting: Experimental and theoretical results, Phys. Rev. Res. \textbf{2},
032029(R) (2020).

\bibitem{exp2} F. Faisant, B. Besga, A. Petrosyan, S. Ciliberto, and S. N.
Majumdar, Optimal mean first-passage time of a Brownian searcher with
resetting in one and two dimensions: experiments, theory and numerical tests,
J. Stat. Mech. \textbf{2021}, 113203 (2021).

\bibitem{anna2} A. S. Bodrova, I. M. Sokolov, Resetting processes with
noninstantaneous return, Phys. Rev. E \textbf{101}, 052130 (2020).

\bibitem{pengbo} P. Xu, T. Zhou, R. Metzler, and W. Deng, Stochastic harmonic
trapping of a L{\'e}vy walk: transport and first-passage dynamics under soft
resetting strategies, New J. Phys \textbf{24}, 033003 (2022).

\bibitem{andrey} W. Wang, A. Cherstvy, H. Kantz, R. Metzler, and I. Sokolov,
Time-averaging and emerging nonergodicity upon resetting of fractional
Brownian motion and heterogeneous diffusion processes, Phys. Rev. E
\textbf{104}, 024105 (2021).

\bibitem{trifce1} T. Sandev, V. Domazetoski, L. Kocarev, R. Metzler, and A.
Chechkin, Heterogeneous diffusion with stochastic resetting, J. Phys. A
\textbf{55}, 074003 (2022).

\bibitem{anna} A. S. Bodrova, A. V. Chechkin, and I. M. Sokolov,
Nonrenewal resetting of scaled Brownian motion,
Phys. Rev. E \textbf{100}, 012119 (2019)

\bibitem{anna1} A. S. Bodrova, A. V. Chechkin, and I. M. Sokolov,
Scaled Brownian motion with renewal resetting,
Phys. Rev. E \textbf{100}, 012120 (2019).

\bibitem{igorjpa} V. P. Shkilev and I. M. Sokolov, Subdiffusive continuous
time random walks with power-law resetting, J. Phys. A, \textbf{55}, 484003
(2022).

\bibitem{anna3} A. S. Bodrova and I. M. Sokolov, Continuous-time random walks
under power-law resetting, Phys. Rev. E \textbf{101}, 062117 (2020).

\bibitem{irina} I. Petreska, L. Pejov, T. Sandev, L. Kocarev, and R. Metzler,
Tuning of the dielectric relaxation and complex susceptibility in a system of
polar molecules: A generalised model based on rotational diffusion with
resetting, Fractal Fract. \textbf{6}, 88 (2022).

\bibitem{gbm} D. Vinod, A. Cherstvy, W. Wang, R. Metzler, and I. Sokolov,
Nonergodicity of reset geometric Brownian motion, Phys. Rev. E \textbf{105},
L012106 (2022).

\bibitem{gbm1} D. Vinod, A. Cherstvy, W. Wang, R. Metzler, and I. Sokolov,
Time-averaging and nonergodicity of reset geometric Brownian motion with drift,
Phys. Rev. E \textbf{106}, 034137 (2022).

\bibitem{trifce} V. Stojkoski, P. Jolakoski, A. Pal, T. Sandev, L. Kocarev,
and R. Metzler, Income inequality and mobility in geometric Brownian motion
with stochastic resetting: theoretical results and empirical evidence of
non-ergodicity, Philos. Trans. A \textbf{380}, 20210157 (2022).

\bibitem{wei} W. Wang, A. Cherstvy, R. Metzler, and I. Sokolov, Restoring
ergodicity of stochastically reset anomalous-diffusion processes,
Phys. Rev. Res. \textbf{4}, 013161 (2022).

\bibitem{net} A. P. Riascos, D. Boyer, P. Herringer, and J. L. Mateos,
Random walks on networks with stochastic resetting, Phys. Rev. E \textbf{101},
062147 (2020).

\bibitem{net1} Y. Ye and H. Chen, Random walks on complex networks under
node-dependent stochastic resetting, J. Stat. Mech. \textbf{2022}, 053201
(2022).

\bibitem{net2} M. Sarkar and S. Gupta, Biased random walk on random networks
in presence of stochastic resetting: exact results, J. Phys. A \textbf{55},
42LT01 (2022).

\bibitem{kiril} K. Zelenkovski, T. Sandev, R. Metzler, L. Kocarev, and L.
Basnarkov, Random walks on networks with centrality-based stochastic resetting,
Entropy \textbf{25}, 293 (2023).

\bibitem{bressloff} P. C. Bressloff, Search processes with stochastic resetting
and multiple targets, Phys. Rev. E \textbf{102}, 022115 (2020).

\bibitem{bressloff1} R. D. Schumm and P. C. Bressloff, Search processes with
stochastic resetting and partially absorbing targets, J. Phys. A \textbf{54},
404004 (2021).

\bibitem{boyer} A. Falc{\'o}n-Corte{\'e}s, D. Boyer, L. Giuggioli, and S. N.
Majumdar, Localization transition induced by learning in random searches,
Phys. Rev. Lett. \textbf{119}, 140603 (2017).

\bibitem{vilk} O. Vilk, D. Campos, V. M{\'e}ndez, E. Lourie, R. Nathan, and
M. Assaf, Phase transition in a non-Markovian animal exploration model with
preferential returns, Phys. Rev. Lett. \textbf{128}, 148301 (2022).

\bibitem{dumas} V. Dumas, F. Guillemin, and P. Robert, A Markovian analysis
of additive-increase, multiplicative-decrease (AIMD) algorithms. Adv. Appl.
Probab. \textbf{34}, 85 (2002).

\bibitem{lopker} A. Lopker, and W. Stadje, Hitting times and the running maximum
of Markovian growth-collapse processes, J. Appl. Prob. \textbf{48}, 295 (2011).

\bibitem{ruinprobs} S. Asmussen, and H. Albrecher, Ruin Probabilities
(World Scientific, Singapore, 2010).

\bibitem{marciniak} E. Marciniak, and Z. Palmowski, On the optimal dividend
problem for insurance risk models with surplus-dependent premiums, J. Optim.
Theor. Appl. \textbf{168}, 723 (2016).

\bibitem{hofstad} R. v. d. Hofstad, S. Kapodistria, Z. Palmowski, and S.
Shneer, Unified approach for solving exit problems for additive-increase and
multiplicative-decrease processes, J. Appl. Probab. \textbf{60}, 85 (2023).

\bibitem{boxma} O. Boxma, D. Perry, W. Stadje, and S. Zacks, A Markovian
growth-collapse model, Adv. Appl. Prob. \textbf{38}, 221 (2006).

\bibitem{boxma1} O. Boxma, D. Perry, and W. Stadje, Peer-to-peer lending: a
growth-collapse model and its steady-state analysis, Math. Meth. Operat. Res.
\textbf{96}, 233 (2022).

\bibitem{lopkertcp} A. Lopker, J. S. H. van Leeuwaarden, and  T. J. Ott, TCP
and iso-stationary transformations, Queu. Syst. \textbf{63}, 459 (2009).

\bibitem{marcus} M. Dahlenburg, A. V. Chechkin, R. Schumer, and R. Metzler,
Stochastic resetting by a random amplitude, Phys. Rev. E \textbf{103}, 052123
(2021).

\bibitem{pierce} J. K. Pierce, An advection-diffusion process with proportional
resetting, E-print arXiv:2204.07215.

\bibitem{shlomi} O. Tal-Friedman, Y. Roichman, and S. Reuveni, Diffusion with
partial resetting, Phys. Rev. E \textbf{106}, 054116 (2022).

\bibitem{hanson} F. B. Hanson and H. C. Tuckwell, Logistic growth with random
density independent disasters, Theoret. Popul. Biol. \textbf{14}, 1 (1981).

\bibitem{gripenberg} G. Gripenberg, A stationary distribution for the growth
of a population subject to random catastrophes, J. Math. Biol. \textbf{17},
371 (1983).

\bibitem{pakes} A. G. Pakes, Limit theorems for the population size of a birth
and death process allowing catastrophes, J. Math. Biol. \textbf{25}, 307
(1987).

\bibitem{brockwell} P. J. Brockwell, J. Gani, and S. I. Resnick, Birth
immigration and catastrophe processes, Adv. Appl. Probab. \textbf{14},
709 (1982).

\bibitem{artalejo} J. R. Artalejo, A. Economou, and M.J. Lopez-Herrero,
Evaluating growth measures in populations subject to binomial and geometric
catastrophes, Math. Biosc. Eng. (MBE) \textbf{4}, 573 (2006).

\bibitem{boyer1} J. Quetzalcoatl Toledo-Marin and D. Boyer, First passage
time and information of a one-dimensional Brownian particle with stochastic
resetting to random positions, E-print arXiv:2206.14387.

\bibitem{kusmierz} {\L}. Ku{\'s}mierz and E. Gudowska-Nowak, Optimal first-arrival
times in L{\'e}vy flights with resetting, Phys. Rev. E \textbf{92}, 052127 (2005).

\bibitem{pccp} R. Metzler, J.-H. Jeon, A. G. Cherstvy, and E. Barkai, Anomalous
diffusion models and their properties: non-stationarity, non-ergodicity, and
ageing at the centenary of single particle tracking, Phys. Chem. Chem. Phys.
\textbf{16}, 24128 (2014).

\bibitem{bouchaud1} J.-P. Bouchaud and A. Georges, Anomalous diffusion in 
disordered media: statistical mechanisms, models and physical applications,
Phys. Rep. \textbf{195}, 127 (1990).

\bibitem{hughes} B. D. Hughes, Random walks and random environments, vol 1:
random walks (Oxford University Press, Oxford, UK, 1995).

\bibitem{sato} K. I. Sato, L{\'e}vy Processes and Infinitely Divisible
Distributions, Cambridge Studies in Advanced Mathematics (Cambridge
University Press, Cambridge UK, 1999).

\bibitem{ourrev} A. Chechkin, R. Metzler, J. Klafter, and V. Gonchar,
Introduction to the Theory of L{\'e}vy Flights. In: R. Klages, G. Radons, and
I. M. Sokolov, editors, Anomalous Transport: Foundations and Applications
(Wiley-VCH, Weinheim, 2008), pp. 129-162.

\bibitem{Linnik} Y. V. Linnik, Selected Transl. Math. Statist. and Prob.
\textbf{3}, 41 (1953).

\bibitem{report} R. Metzler and J. Klafter, The random walk's guide to anomalous
diffusion: A fractional dynamics approach, Phys. Rep. \textbf{339}, 1 (2000).

\bibitem{special_functions} G. Andrews, R. Askey, and R. Roy, Special Functions
(Cambridge University Press, Cambridge UK, 1999).

\bibitem{ahlfors} L. Ahlfors, Complex Analysis (McGraw-Hill, New York, 1979).

\bibitem{pantograph} L. Fox, D. F. Mayers, J. R. Ockendon, and A. B. Tayler, On
a functional differential equation, J. Inst. Math. Appl. \textbf{8}, 271 (1971).

\bibitem{mathai} A. M. Mathai and R. K. Saxena, The H-Function with
applications in statistics and other disciplines (Wiley Eastern, New
Delhi, 1978).

\bibitem{walterg} W. G. Gl{\"o}ckle and T. F. Nonnenmacher,
Fox-function representation of non-Debye relaxation processes,
J. Stat. Phys. \textbf{71}, 741 (1993).

\end{thebibliography}
\end{document}